# Beam Loss Monitors

*Kay Wittenburg*
DESY, Hamburg, Germany

**Abstract**
This lecture covers the fundamental aspects of the measurement of beam losses including their use for beam diagnostic and safety issues. The detailed functionality and detection principle of various common beam loss monitors are also presented, with a focus on their intrinsic sensitivity.

## 1 Introduction

This lecture based on my "Beam Loss Monitors" report from the Beam Instrumentation CAS in Tuusula, Finland, 2018. I have added some more recent comments and references to it.

The important characteristic in "Beam Losses" is the number of particles lost per unit time $\Delta N/\Delta t$. The beam lifetime $\tau$ is defined with $N(t) = N_0\ e^{-t/\tau}$ where $N_0$ is the initial intensity. At the accelerator design stage, such questions as, "What is the minimum/maximum possible beam lifetime?" and, "What (and where) is the tolerable beam loss rate for accelerator components?" have to be answered. Beam loss monitor systems are typically designed for measuring the amount and location of beam losses along an accelerator or storage ring. An appropriate design of such a system depends on various aspects, in particular:

- What kinds of beam losses are to be detected?
- A detailed understanding of the beam loss mechanism and the adjacent processes which make a beam loss detectable by the loss monitors.
- The type of beam loss monitor.

This lecture is divided into three parts, reflecting the aspects from above.

Chap. 2: Loss classes: Different kinds of beam losses are discussed to give a feeling for the wide range of applications of beam loss monitors and to motivate the following two topics.

Chap. 3: Principles of loss detection: The fundamental aspects of measuring beam losses are discussed.

Chap. 4: Beam Loss Monitors (BLMs): Some typical BLMs are discussed with the focus on their sensitivity.

Since each topic could be a lecture on its own, this report is supported by many references which go into more depth on these topics.

## 2 Loss classes

In this section I would like to discuss some examples of interesting measurements and applications for BLMs.

Beam losses are divided into two different classes, namely irregular losses and regular losses [1]. The specification for acceptable beam losses as a function of beam momentum (see Section 2.1) and the minimum required sensitivity defines the required dynamic range for the BLMs. Additional sensitivity combined with a larger dynamic range extends the utility of BLM systems for diagnostic work (Section 2.2).

## 2.1 Irregular beam losses

This class of losses (sometimes also called uncontrolled or fast losses) contains *avoidable (!) losses* of higher levels which are often the result of a misaligned beam or of a faulty condition of accelerator elements. A typical example is a trip of the RF which, in electron storage rings, causes fast beam loss as a result of the energy loss of the electrons; vacuum problems also fall into this category. These losses may be distributed around the machine. A well-designed collimator system might capture most of the losses, but even then, a fraction of the losses may cause problems at other locations. These losses should be avoided and should be kept to low levels:

- to keep activation low enough for hands-on maintenance, personal safety and environmental protection.
- to protect machine parts from beam related (radiation) damage (incl. quench protection and protection of the detector components)
- to achieve long beam lifetimes/efficient beam transport to get high integrated luminosity or stable synchrotron light brilliance for the related experiments.

Sometimes such losses have to be tolerated even at a high level at low repetition rates during machine studies. A beam loss monitor system should define the allowed level of those losses. The better protection is against these losses, the less likely is down time due to machine damage. Very often the BLMs are part of the machine protection system [2] to keep those losses below a certain level (defined by the BLM system) in order to avoid beam-loss-related 'damage' of components. A post mortem event analysis is most helpful to understand and analyse the faulty conditions.

### *2.1.1 Some examples of irregular beam losses*

#### *2.1.1.1 Superconducting machines: Quench protection*

Superconducting accelerators need a dedicated BLM system to prevent beam-loss-induced quenches. Such a system has to detect losses fast enough before they lead to a high energy deposition in the superconducting material which might lead to a quench. Examples of such systems can be found at HERA, Tevatron and at LHC [2], [3]. Typically, a BLM system implements several individual thresholds for the amount of beam losses, and automatically dumps the beam if one or more of them are exceeded. The thresholds can depend on various conditions, e.g., the beam energy, the BLM position, the allowed heat deposition in the superconductor. Figure 1 shows an example of a post mortem archive of a beam loss in the HERA superconducting proton ring. To prevent unwanted beam dumps due to faulty conditions of one BLM, it might be useful to add another threshold for how many BLMs have to be above their individual threshold before a dump is triggered. In this case one might have to multiply the amount of BLMs at critical locations by that number to ensure safe conditions even with very localized beam losses or at very sensitive locations (e.g., high lattice dispersion in superconducting magnets).

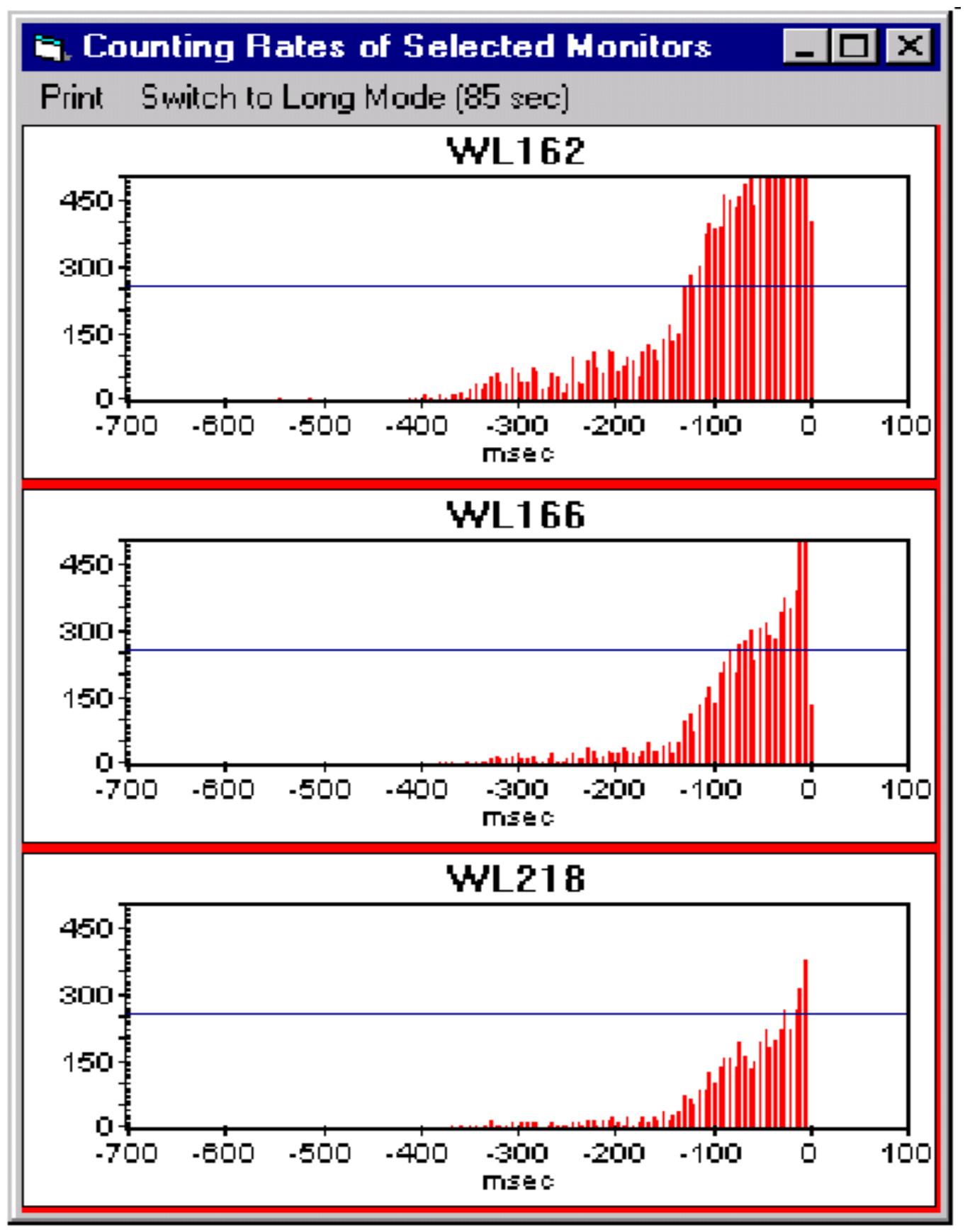

**Fig. 1:** Beam losses in HERAp. Each vertical line corresponds to a beam loss rate within 5.2 ms. The horizontal line indicates the individual threshold of each BLM (three BLMs at different locations are shown). After four BLMs reached their individual threshold (horizontal line) the beam was dumped by the system in order to protect the magnets from quenching. The history of the last 700 ms before the dump and 100 ms after the dump are shown.

#### *2.1.1.2 Activation of environment due to losses*

The activation of components should be kept low enough for hands-on maintenance [typically not more than 0.1 rad/h residual activation (30 cm from surface, after 4 h cool-down)], personal safety, and environmental protection. As a rule of thumb this value means for a hadron accelerator an allowed slow beam loss rate of about 1 W/m, or for a 1 GeV hadron accelerator an allowed loss rate of about 13 particles/m at 500 MHz. For electron accelerators the rates might be higher since activation is less likely with interactions of electrons with matter. Figure 2 shows as an example the strong correlation between measured losses and residual activation at KEK. But also in electron machines, the radiation damage of undulators in synchrotron light sources and FEL became a serious problem [4]. To avoid activation in general and damage of sensitive components the loss rates should be kept as low as possible with the help of a BLM system.

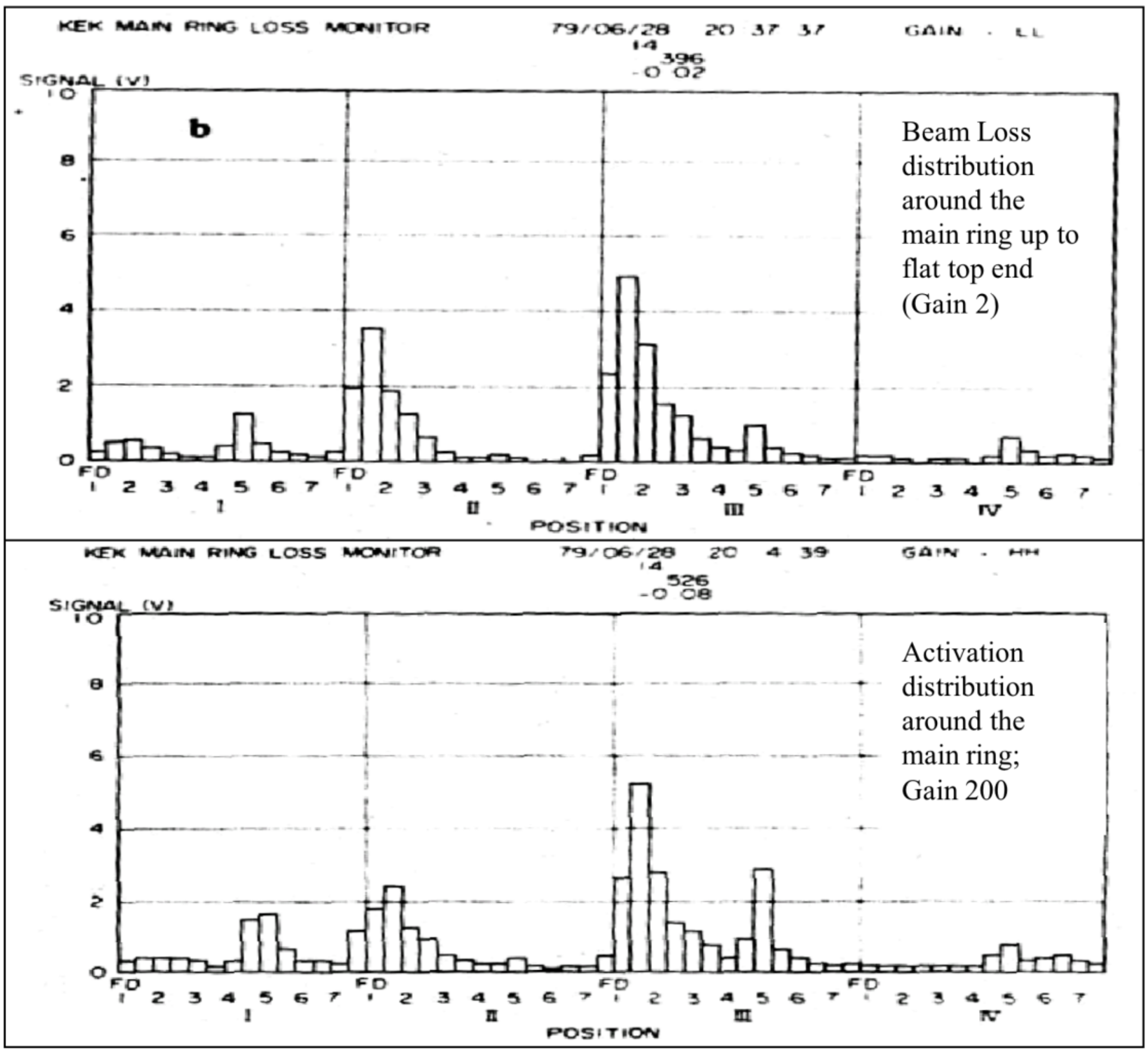

**Fig. 2:** Correlation between beam losses (upper part) and activation (lower part) at KEK measured by beam loss monitors [5]

### *2.1.1.3 Commissioning: Obstacles*

During commissioning of an accelerator, a BLM system can be most helpful in detecting several problems which might prevent circulating beam. Typical examples are badly connected RF fingers which point into the beam pipe and act as a target for the beam (not to mention the empty(!) beer bottles in the LEP beam pipe). The readings of the BLMs around the ring showed such an obstacle during the commissioning of the RHIC ring [6], see Fig. 3. This made the localization of the problem much easier, even a steering of the beam with a local bump around this obstacle immediately provided circulating beam.

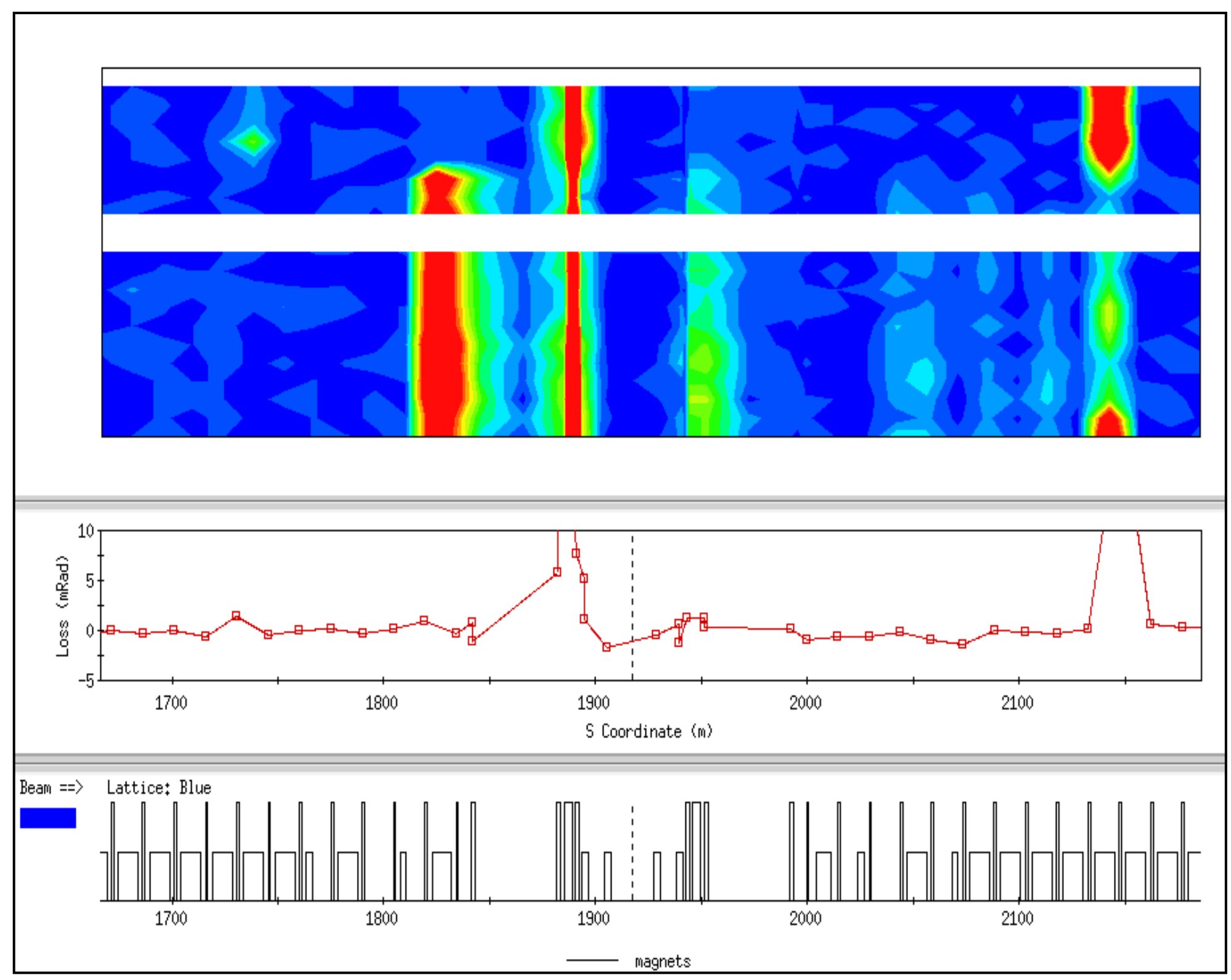

**Fig. 3:** Loss pattern evolution in RHIC as beam was steered locally around an apparent obstacle at s $\simeq$ 1820 m. When the losses there went away, beam began circulating for thousands of turns. From Ref. [6].

### *2.1.1.4 Vacuum problems (Coulomb scattering)*

BLMs can help to detect even very small vacuum leaks. Beam particles scatter on the residual gas, so a pressure bump can be seen from a higher loss rate in that area. Figure 4 shows an example from HERAe, where the ring-wide BLM system detected a higher loss rate at two locations, which were correlated with two small vacuum leaks. After the repair, the lifetime $\tau$ of the beam immediately increased to the original value.

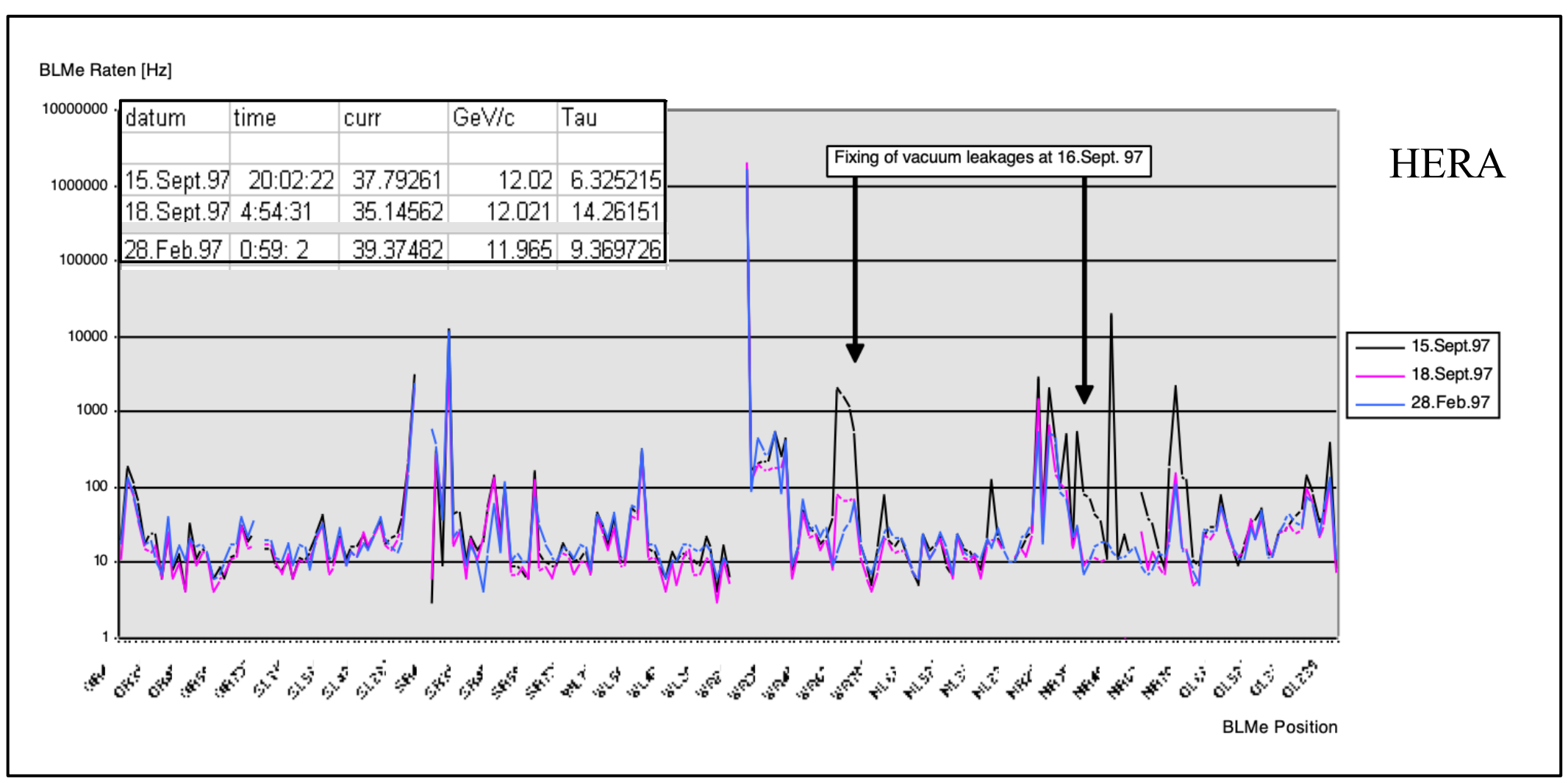

| datum | time | curr | GeV/c | Tau |
|---|---|---|---|---|
| 15.Sept.97 | 20:02:22 | 37.79261 | 12.02 | 6.325215 |
| 18.Sept.97 | 4:54:31 | 35.14562 | 12.021 | 14.26151 |
| 28.Feb.97 | 0:59: 2 | 39.37482 | 11.965 | 9.369726 |

**Fig. 4:** Two vacuum leaks detected by the BLM measurements at HERAe in September 1997 (arrows). The BLM rates are plotted on the vertical scale, on the horizontal axis are the BLM locations around the whole circumference of the ring. The leak was detected on 15 September 1997 and fixed on 16 September 1997. On 18 September 1997 the loss rates were back to the original rates measured on 28 February 1997. The two gaps in the plot and the adjacent huge loss rates are at the locations of the two collision points where the interaction products of the two colliding beams create the loss signals.

### *2.1.1.5 Microparticles and UFOs*

Unexpected reversible and irreversible jumps in the lifetime were observed when operating HERAe with electrons [7] (Fig. 5) and LHC with protons (Unidentified Falling Object (UFO)) [8]. The BLMs were most helpful for a better understanding of this problem, they localized very precisely the origin of the jumps. It turned out to be microparticles emitted from ion getter pumps (HERA) and dust particles falling into the beam (LHC).

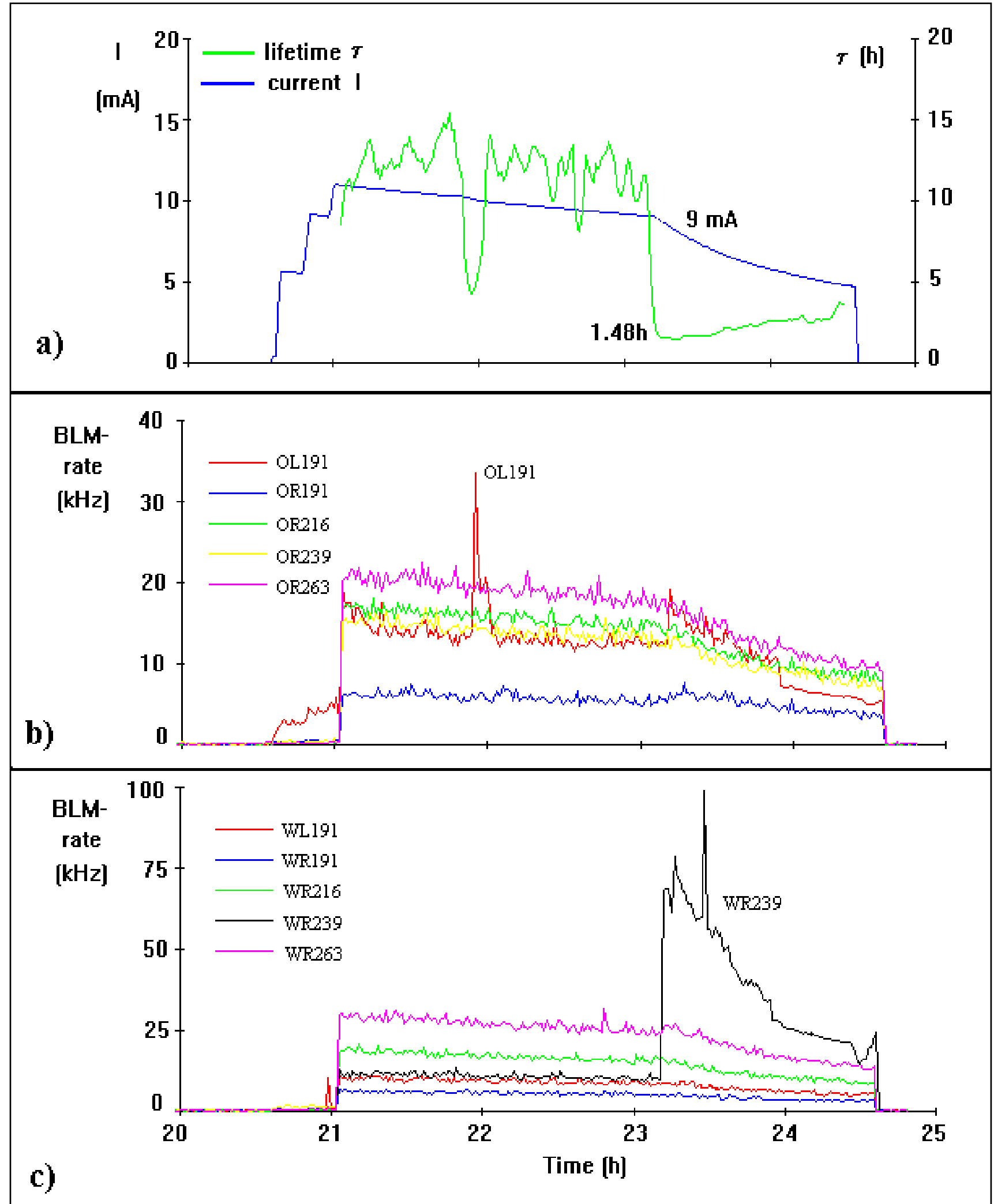

**Fig. 5:** Lifetime reduction events correlate well with losses seen in the HERA electron loss monitors. In this example a brief disruption of lifetime is seen in the loss monitor OL191, and an irreversible disruption is seen in the monitor WR239. From Ref. [7].

### *2.1.1.6 High current and high brilliance machines (ring or linac): Destruction of vacuum components*

A serious problem for high current and high brilliance accelerators is the high-power density of the beam. A misaligned beam is able to destroy the beam pipe or collimators and may break the vacuum (Fig. 6). This fact makes the BLM system one of the primary diagnostic tools for beam tuning and equipment protection in these machines.

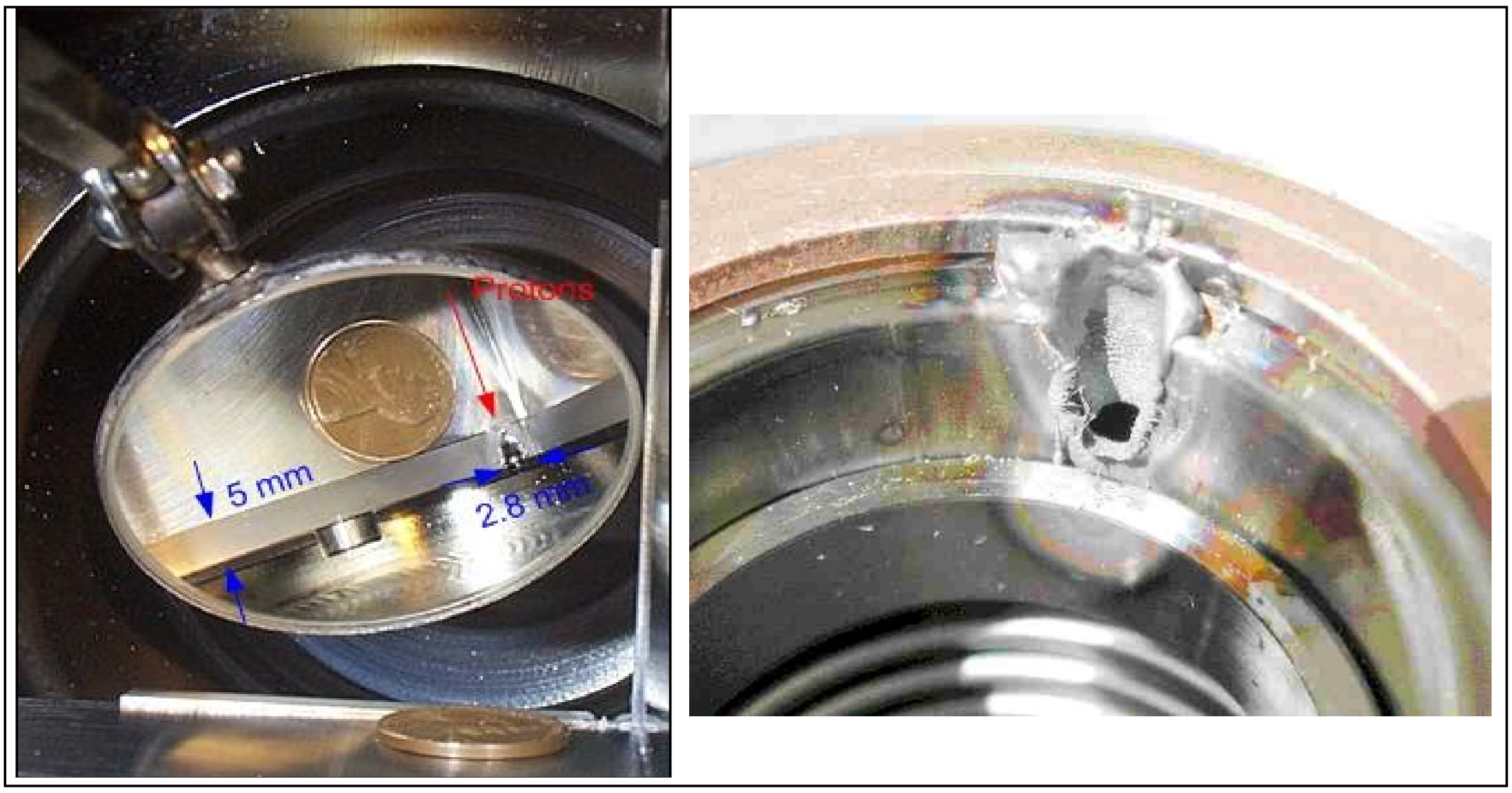

**Fig. 6:** Both figures show serious damage to vacuum components in the Tevatron [9] (left) and ELBE [10] (right) due to direct beam impact. In both cases no BLM system was available at the time.

### 2.2 Regular beam losses

Regular losses (sometimes also called controlled or slow losses) are typically un-avoidable and are localized at the collimator system or at other (hopefully known) aperture limitations. They might occur continuously during operation and correspond to the lifetime/transport efficiency of the beam in the accelerator. The lowest possible loss rate is defined by the theoretical beam lifetime limitation due to various effects, e.g., Touschek effect, beam–beam interactions, collisions, transversal and longitudinal diffusion, residual gas scattering, halo scraping, instabilities. These losses are suitable for machine diagnostics with a BLM system; the system should be sensitive enough to enable machine fine tuning and machine studies with the help of BLM signals:

- sometimes even at low beam intensity to avoid high losses and/or
- during machine commissioning and
- at various energies during acceleration.

It is clearly advantageous to design a BLM System which is able to deal with both (irregular and regular) loss modes; therefore, a very high dynamic range is required.

#### *2.2.1 Some examples of regular beam losses*

##### *2.2.1.1 Injection studies*

BLMs are useful to improve the injection efficiency, even at low injection current (radiation safety issue). They are more sensitive than current transformers and they can distinguish between transversal mismatch (high losses at high beta values) and energy mismatch (high losses at high dispersion values) and unpredicted small apertures (losses somewhere else).

Figure 7 shows an example from the ALS. Several BLMs report high count rates at injection. From this graph one can identify the sites of the highest beam loss. Several peaks result from different centres of beam loss. An online optimization of the injection elements is possible by minimizing these loss rates.

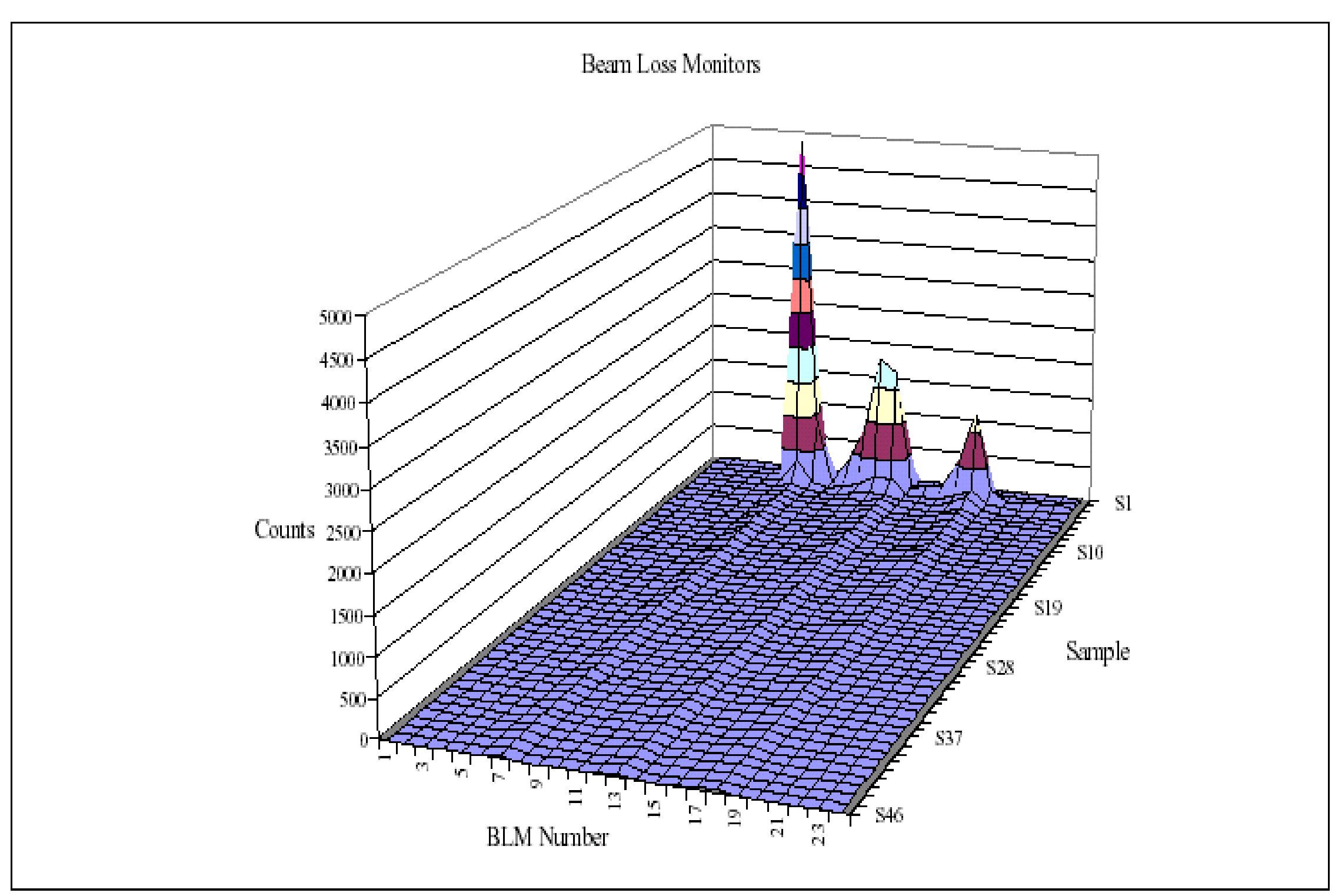

**Fig. 7:** Surface plot of beam loss at injection, from J. Hinkson, "ALS beam instrumentation: beam loss monitoring," February 1999, not published.

### *2.2.1.2 Lifetime limitations (Touschek effect, etc.)*

In addition to faulty conditions, there are unavoidable effects which limit the beam lifetime in an accelerator, e.g., vacuum lifetime (Coulomb scattering), Touschek effect, quantum lifetime.

Preferred BLM locations for the detection of Touschek scattered (beam-) electrons (or positrons) are high dispersion sections which follow sections where a high beam density was reached. Since the two colliding electrons lose and gain an equal amount of momentum, they will hit the inside and outside walls of the vacuum chamber. In contrast, elastically or inelastically scattered electrons [e.g. on residual gas atoms (Coulomb) or on photons (Compton) or high-energy synchrotron radiation photons (Quantum)] are lost at aperture limits. If the energy carried away from the beam electron is too large, the electron gets lost after the following bending magnet (or at high dispersion) only on the inside wall of the vacuum chamber. A BLM system with good selectivity to the different loss mechanisms is a very useful tool for various kinds of beam diagnostics, especially in Touschek limited accelerators; the Touschek loss rate depends on the 3-dimensional electron density and on the spin of the scattering particles. Therefore, any change in one or more of these parameters has an influence on the loss rates at the selected monitors. An example is shown in Fig. 8 where the BLM system at BESSY was used to determine the (desired) vertical beam blow-up due to a resonant head–tail mode excitation [11]. BLM systems were also used to calibrate precisely the beam energy and observe its variation in time by using resonant depolarization of the beam [12].

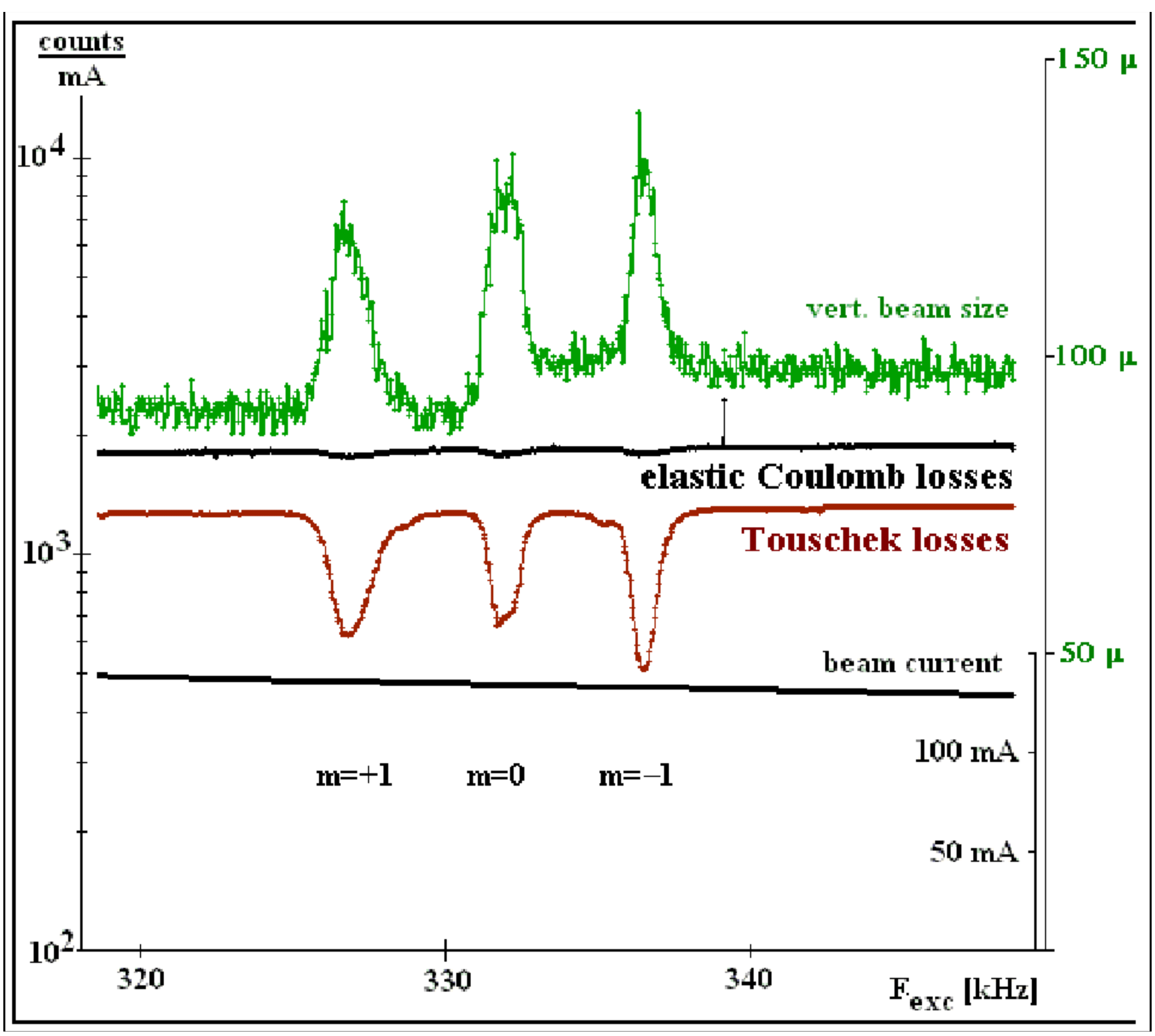

**Fig. 8:** Vertical beam size, Touschek and Coulomb loss rates during excitation of a vertical head–tail mode in BESSY [11]. Clearly observable is the decrease of the Touschek loss rate (BLMs at the outside of the ring) with increasing vertical beam size, while the Coulomb losses stay nearly constant (BLMs on the inside of the ring).

### *2.2.1.3 Tail measurements*

Non-Gaussian tails in the beam distribution produce lower beam lifetimes and background in experimental detectors. These tails are difficult to detect with standard beam profile monitors because of their small population with respect to the core of the beam [13]. A combination of scrapers and BLMs is a very sensitive way to measure tail populations. The measurement and scraping can be done by moving the scraper closer to the beam core in small steps, measuring at each step the response of the adjacent BLM. The movement of the scraper can be adjusted according to the loss rate to avoid scraping the beam core. Reference [14] shows measurements at LEP where the dominant processes of tail creation were Compton scattering on thermal photons (horizontal) and beam–beam bremsstrahlung (vertical).

In high-energy proton accelerators, lifetime limitations may arise from proton diffusion due to beam–beam interactions and tune modulation due to ground motion. The ground motion frequencies and diffusion parameters can be measured with BLMs at the collimators [15], [16]. Figure 9 shows BLM measurements at HERAp where the frequency spectrum of the measured losses of a very stable beam corresponds very well to that of the ground motion.

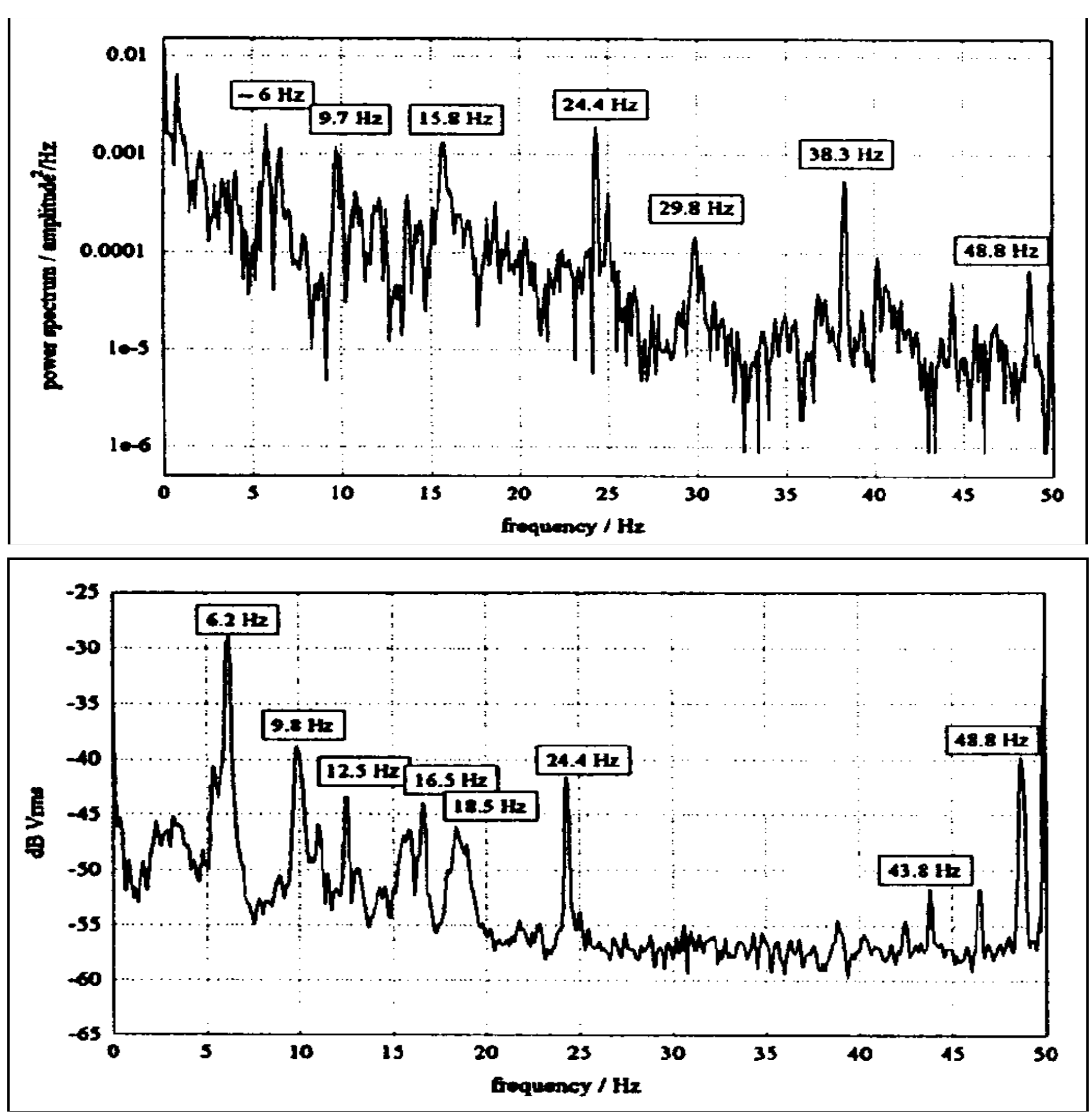

**Fig. 9:** Frequency spectrum of ground motion (upper) and frequency spectrum of losses at a collimator (lower), from Ref. [16].

### *2.2.1.4 Tune scans*

Regular losses can also be used for optimizing the machine lattice by studying the tune dependence of the losses [11]. A tune scan is also useful for studying nonlinear resonances caused by insertion devices [17]. Figure 10 shows the scanning of the various tunes in both planes and their corresponding beam loss rates. In this example a $Q_x + 3Q_y$ resonance was clearly identified which causes harmful beam losses at a certain location (equipped with a BLM) in BESSY. The localization of a certain loss is clearly an advantage with respect to a beam lifetime measurement which does not provide position information. With the help of BLMs tune scans can be made faster because the sensitive direct loss measurement is faster than a precise lifetime determination.

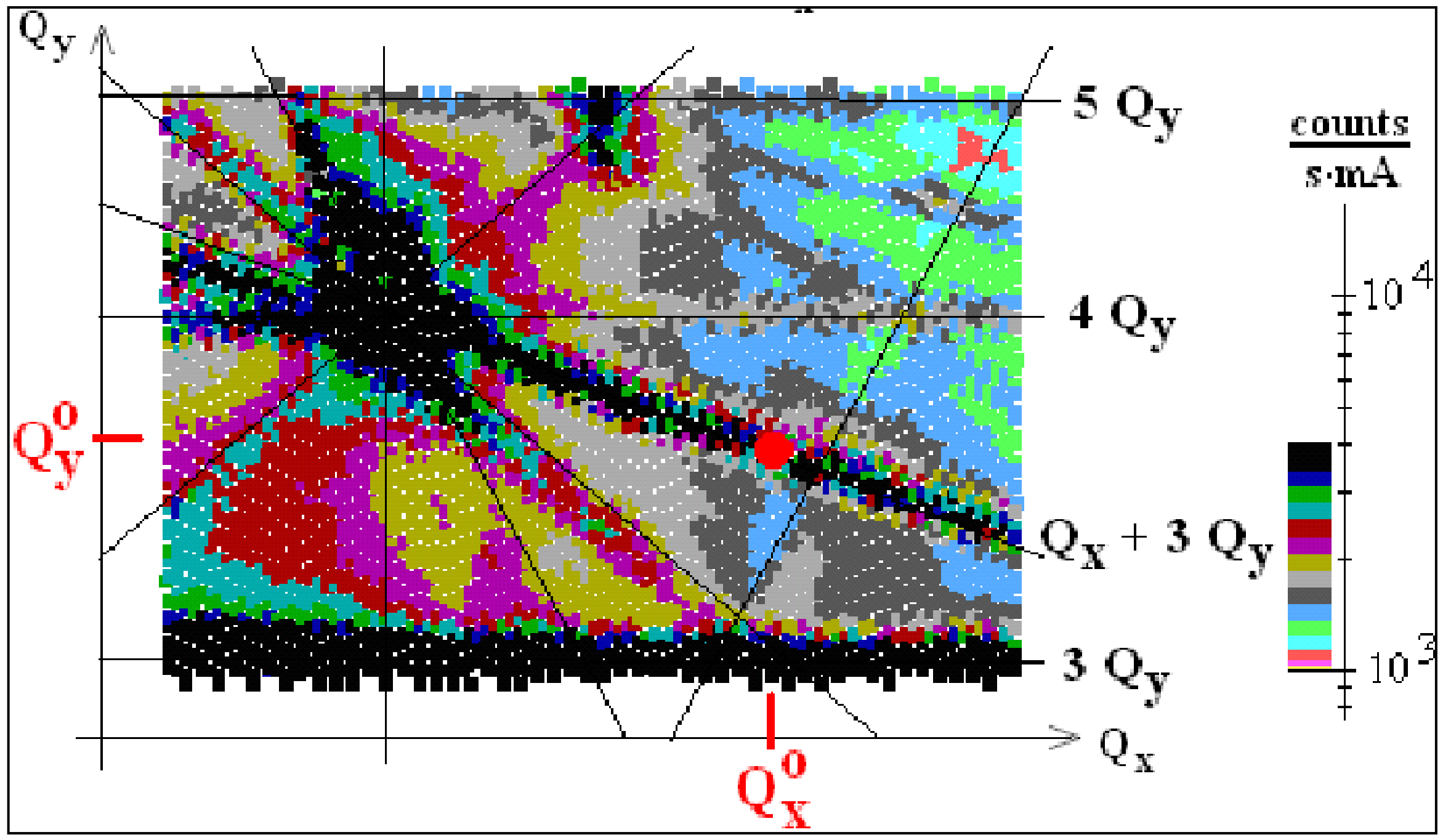

**Fig. 10:** Two-dimensional tune scan at BESSY [11]

## 3 Principles of loss detection

This section describes some common aspects of beam loss measurements. Note that systems like differential beam current measurements (e.g. [18]) and dose measurements (or activation) (e.g. [19]) are not the subject of this session. They have a very rough position resolution or a very long time constant, respectively. This session concentrates on dedicated beam loss measurements with some numbers of BLMs and a time resolution in the order of some turns or even faster, up to bunch-by-bunch resolution. BLMs at special locations may serve also for dedicated beam studies as explained in the previous section.

### 3.1 Considerations in selecting a beam loss monitor

R.E. Shafer gave a very good tutorial about beam loss monitoring at the 2002 Beam Instrumentation Workshop [20] where he summarized the important factors for selecting the right type of BLM for a specific application. Consideration of these parameters gives a good guide for finding (or designing) the best monitor type for a particular beam loss application:

- *Intrinsic sensitivity*
- Type of output (current or pulse)
- Ease of calibration (online)
- System end-to-end online tests
- Uniformity of calibration (unit to unit)
- Calibration drifts due to aging, radiation damage, out-gassing, etc.
- *Radiation hardness (material)*
- Reliability, Availability, Maintainability, Inspect ability, Robustness
- Cost (incl. Electronics)
- Shieldability from unwanted radiation (Synchrotron Radiation)
- *Physical size*
- Spatial uniformity of coverage (e.g. in long tunnel, directionality)
- Dynamic range (rads/sec and rads)
- *Bandwidth (temporal resolution)*
- Response to low duty cycle (pulsed) radiation

- Instantaneous dynamic range (vs. switched gain dynamic range)
- Response to excessively high radiation levels (graceful degradation)
- *Signal source*
- *Positioning.*

In the discussion of some BLM types in Section 4, I shall stress the topics in italics a little bit more while the last two topics are discussed in Sections 3.2 and 3.3.

### 3.2 The signal source or what should a beam loss monitor monitor?

In case of a beam loss, a BLM has to establish the number of lost particles in a certain position and time interval. A typical BLM is mounted outside of the vacuum chamber, so that the monitor normally observes the shower caused by the lost particles interacting in the vacuum chamber walls or in the material of the magnets. The number of detected particles (amount of radiation, dose) and the signal from the BLM should be proportional to the number of lost particles. This proportionality (or efficiency ε) in terms of signal/lost particle depends on the type of BLM (its intrinsic sensitivity $S_{BLM}$), the position of the BLM with respect to the beam, the type of the lost particles, and the intervening material. But also, on the momentum of the lost particles, which may vary by a large ratio during the acceleration cycle. It is nearly impossible to *measure* the exact efficiency of BLMs around an accelerator because of the impossibility of generating beam losses at exactly one location in an accelerator since a significant amount of beam particles will be scattered and lost at other places at the same time. Therefore the usual way of defining the overall efficiency ε is to use Monte Carlo programs like EGS (for electrons), MCNPX, MARS, and SHIELD, FLUKA or GEANT4 (for Hadrons) [21]. With the help of these programs one can simulate the particle loss at a certain location together with the whole geometry (beam pipe, magnets, valves) near the BLM, the corresponding magnet and electrical fields of accelerator components and the BLM itself (see Section 3.3). The amount of beam loss at a certain location is then defined as (see Section 3.3)

$$\text{Number of lost particles } N = Signal/\varepsilon.$$

The signal of a BLM depends on the type of BLM and is typically a current or a count rate, in some cases it might be 'light' which is then converted to current by photosensitive elements like photomultipliers.

The common signal source of beam loss monitors is mainly the ionizing capability of the charged shower particles crossing the BLM. Their ionization loss is described by the Bethe–Bloch formula:

$$-\frac{dE}{dx} = \frac{4\pi}{m_e c^2} \cdot \frac{n Z^2}{\beta^2} \cdot \left(\frac{e^2}{4\pi \varepsilon_0}\right)^2 \cdot \left[\ln\left(\frac{2 m_e c^2 \beta^2}{I \cdot (1-\beta^2)}\right) - \beta^2\right] \tag{1}$$

With $\beta = v/c$ and $I \approx 16 \cdot Z_T^{0.9} eV$. I is the mean excitation energy of the target material of density ρ, atomic number $Z_T$ and atomic mass $A_T$. Z is the projectile charge and n is the electron density of the target material.

There is a region at $\beta\gamma = p/m_0c \approx 2$ where dE/dx reaches a minimum (see Fig. 11). A particle having a mean ionizing energy loss in the region of the minimum ionization is often called minimum ionizing particles or MIP. The value of the energy deposited by a MIP is $dE/dx_{MIP} \approx 1\text{-}2\ MeV/(g/cm^2)$ and is valid for many materials. Above this point there is a logarithmic increase of dE/dx with energy, caused by relativistic effects. This increase is considered in the following to be negligible, so that MIP means a particle with $\beta\gamma = p/m_0c \geq 2$.

The energy from such an ionization loss can be used to create electron/ion pairs (current) or photons in the BLM detector material. An energy deposition in (detector-) material can be given in units of rad or Gy (=100 rad). Using the definition of 1 rad = 100 ergs/gram and 1 MeV = $1.6\cdot10^{-6}$ ergs leads to another definition, in terms of MIPs/cm$^2$.

$$1\ rad = \frac{100\ ergs}{gram} \cdot \frac{1\,MeV}{1.6\cdot10^{-6}\ ergs} \cdot \frac{MIP\ gram}{2\,MeV\ cm^2} = 3.1\cdot10^7\ \frac{MIP}{cm^2} \qquad (2)$$

(From Ref. [20])

The intrinsic sensitivity $S_{BLM}$ of a beam loss monitor can now be described in terms of either energy deposition (rad) in a BLM, or in terms of a charged particle (MIPs) flux ($3.1\cdot10^7$ MIPs/cm$^2$) through a BLM (Chap. 3). But note that this relation is true only for a radiation field which contains MIPs only! This is nearly true for high energy accelerators with the BLM close to the beam pipe. As pointed out before, a careful study by Monte-Carlo simulations is recommended to get the real energy- and particle- spectrum of the radiation.

Another typical signal source of BLMs is the “secondary emission” from (detector-) surfaces. When a charged particle crosses a metallic surface, secondary electrons are emitted from it. The Sternglass formula describes the secondary emission coefficient of various materials in terms of “number of emitted electrons” for primary electron, proton and ion impacts [22], [23]. This coefficient is for many metals in the order of a few %/MIP (see Fig. 12).

Sometimes also Bremsstrahlung is used as a signal source, see e.g. [24], [25], but this method is not discussed here because of its very special application.

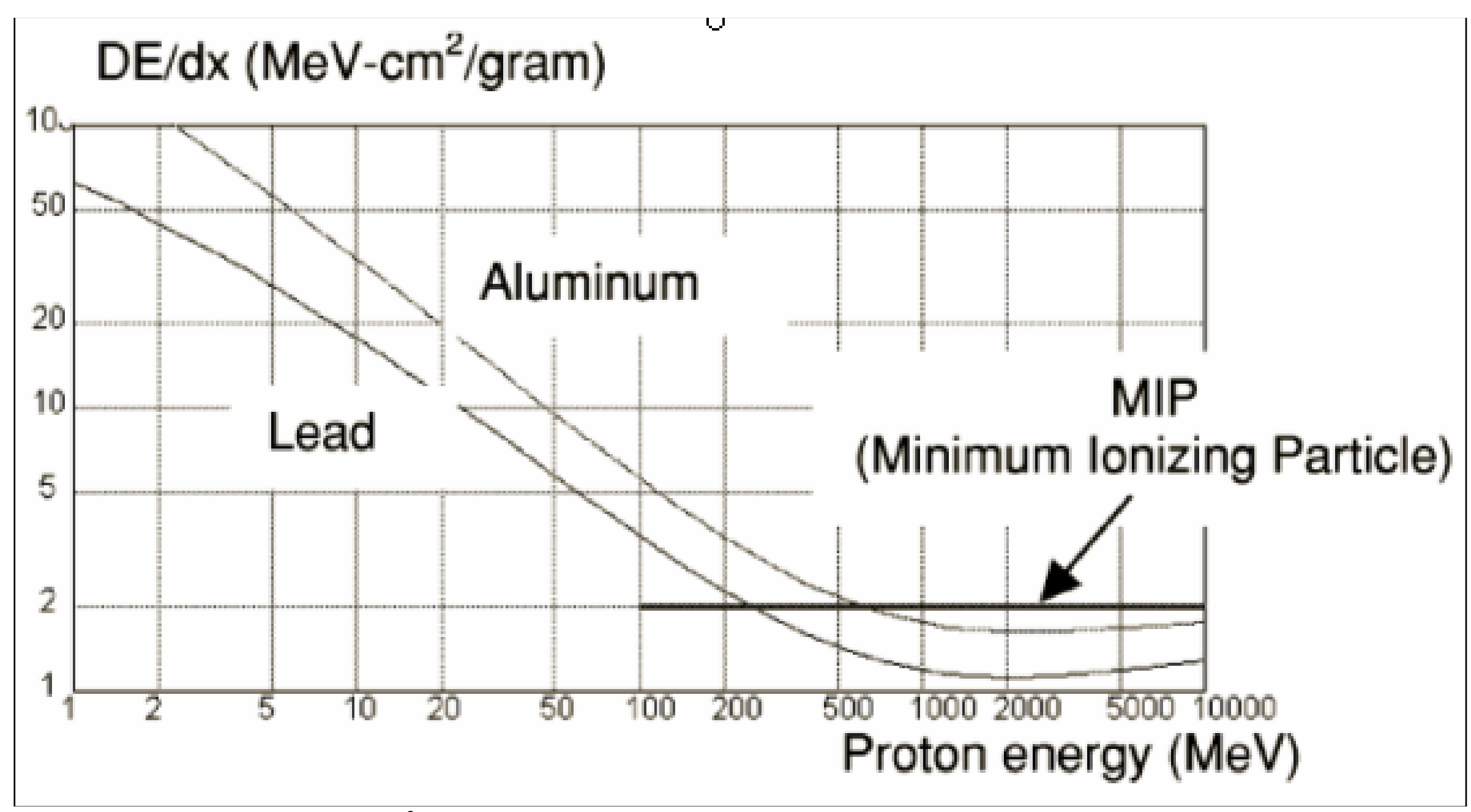

**Fig. 11:** The *dE/dx* [MeV/(g/cm$^2$)] for protons in lead vs. proton energy [MeV/c] (from [20]). Note that e.g. an electron with an energy of *≥1 MeV/c (≈ 2 · p/m$_0$c)* is also considered as a MIP with *dE/dx* ≈ 1-2 MeV/(g/cm$^2$) for many materials.

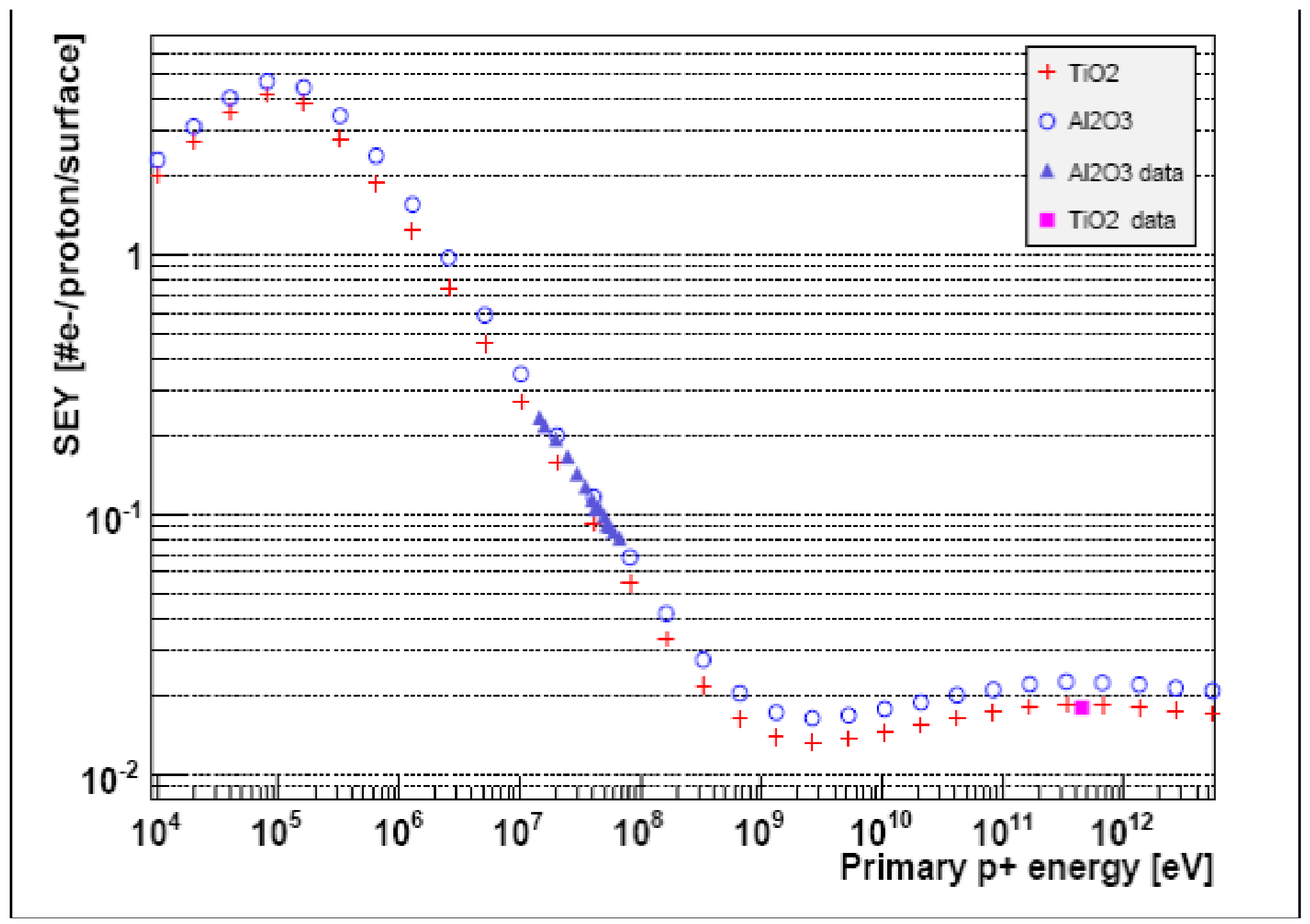

**Fig. 12:** Secondary emission coefficient vs. primary particle energy (here protons), from Ref. [26]

### 3.3 Positioning of beam loss monitors

Each BLM Signal ($Sig_{BLM}$) at each location needs its own specific efficiency calibration ε in terms of signal/lost particle, $Sig_{BLM}(t)/N_{loss}(t)$, especially while the BLM detects secondaries and not the primary lost particle. This calibration can be calculated by two steps:

1. Using a Monte Carlo program using the (more or less) exact geometry and materials between the beam and the BLM to understand the measured *Loss Efficiency* $\varepsilon_{BLM}$

$$\varepsilon_{BLM} = N_{BLM}(t) / N_{loss}(t)$$

with $N_{BLM}(t)$ = rate of (signal-generating) particles reaching the BLM.

2. Knowing the *Sensitivity* of the BLM ($S_{BLM}$) on (signal-generating) particles.

$$Sig_{BLM}(t) = S_{BLM} * N_{BLM}(t) \Leftrightarrow \quad N_{loss}(t) = Sig_{BLM}(t) / (\varepsilon_{BLM} * S_{BLM}).$$

The loss of a high-energy particle in the wall of a beam pipe results in a shower of particles which leak out of the pipe (low-energy beam particles which do not create a shower leakage outside the vacuum pipe wall will hardly be detectable by a loss monitor system). The efficiency of a loss detector will be highest if it is located at the maximum of the shower. Examples can be found in Refs. [27]-[30]. Monte Carlo simulations were used to find the optimum locations for the monitors, as well as to calibrate the monitors in terms of 'lost particles/signal'. The intensity and the length of the shower depends on the energy of the lost particle and ranges from some metres for very high proton energies [28], [29], [31] and neutron detection [32]-[34] to a few centimetres for medium electron energies [35], [36]. Therefore, the expected location of lost particles has to be studied in advance so as to situate the monitors at the right place, especially lower energy accelerators where the length of the shower is quite small. This implies that an understanding of the loss mechanism and dynamics in the accelerator are necessary to predict the positions of losses. For example, Refs. [29], [35], [37] report detailed particle tracking studies being done to follow the trajectory of an electron in the accelerator after an energy loss due to scattering on a residual gas molecule or on a microparticle. There are many different reasons for beam losses and a complete beam loss system has to be carefully designed for the detection and analysis of a specific loss mechanism (see examples in Section 2).

Figure 13 shows an example of beam losses due to inelastic scattering (on residual gas or microparticles). This example also shows that a distortion of the beam results in beam offsets which are largest in the middle of focusing quadrupoles. This is also true for the beam size, which is large at high β functions. Therefore, the preferred locations for beam losses (and therefore for BLMs) might be aperture limits like collimators, scrapers, locations with high dispersion or high β functions. Since the quadrupoles always have a local β maximum, either in $x$ or in $y$, they are the preferred locations for BLMs around an accelerator. This is especially true for BLMs which have to serve the quench protection system in superconducting accelerators [2], [3], [4], [27]. Again, detailed Monte Carlo simulations can help find the optimum positions of BLMs at the superconducting quadrupoles as well as the efficiencies at these positions.

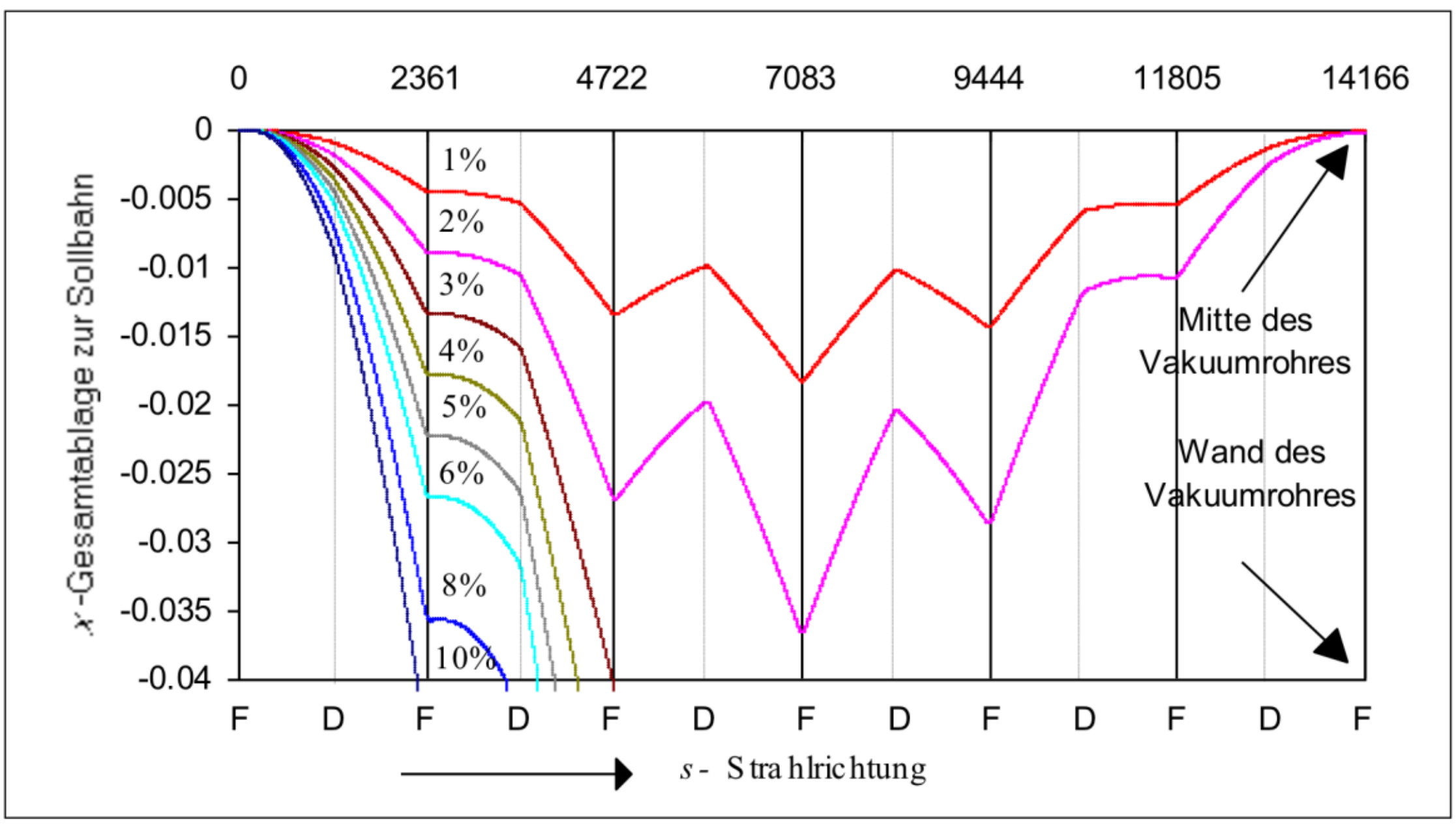

**Fig. 13:** The beam-electrons lose energy DE of between 1% and 10% due to inelastic scattering (bremsstrahlung) mainly on the nuclei of the residual gas molecules. The deviation of the electron orbit from the nominal orbit x along some FODO cells s depends on the dispersion function in the accelerator and on DE. Therefore, the electrons may be lost at their aperture limit behind the following bending magnet on the inside wall of the vacuum chamber at 4 cm [37].

## 4 Types of beam loss monitors

In the following sections I discuss some typical BLM detectors with the emphasis on their intrinsic sensitivity $S_{BLM}$, but other aspects including radiation hardness, size, and temporal resolution are also treated. Additional overviews of BLMs and detectors can be found, for example, in Refs. [38], [39] and in various text books [40], [41].

### 4.1 Short ionization chamber

Short ionization chambers are used as BLMs in many accelerators. An ionization chamber in its simplest form consists of two parallel metallic electrodes (anode and cathode) separated by a gap of width $D$.

The gap is filled with gas (air, argon, xenon[1]) or liquid of density ρ and defines the sensitive volume of the chamber. High voltages, *V*, up to several kV, are applied between anode and cathode. Ionizing particles traversing the sensitive volume ionize the gas or liquid and produce electron–ion pairs. The electric field $E = V/D$ causes electrons and positive ions to drift in opposite directions towards the anode and cathode, respectively. The number of electrons reaching the anode depends on the applied voltage. If the voltage is very small, the electrons produced by ionization recombine with their parent ion. The electrons can escape this initial recombination if the electric field is larger than the Coulomb field in the vicinity of the parent ion. The number of electrons collected at the anode increases with voltage up to saturation when all charges are collected. In this *ionization region* the created charge should not depend on the applied voltage. At higher voltages, the electrons gain enough energy to produce ionization on their path which is called the *proportional region* and the *Geiger–Müller region* is where saturation is reached (see Fig. 14).

The flatness of the plateau of the ionization region depends on the collection efficiency of the electrons or ions on the electrodes. Especially at high radiation levels, electrons on their way to the anode may be captured by positive ions produced close to their trajectory (by other incoming particles) and do not contribute to the charge collection. Therefore, high voltage and a small gap *D* are preferred to achieve a high dynamic range of an ionization chamber. This effect was well documented during the improvement of the ionization chamber BLM for SNS [42]. A smaller gap has an additional effect on the response time of a chamber: the response time is defined by the transit time t of the created charge through the gap (here for ions) [43]:

$$t = D/v_{ion} = D^2/[\mu_0 V (P_0/P)]$$

with the ion speed:

$$v_{ion} = \mu_0 E(P_0/P) = \mu_0 (V/D)(P_0/P),$$

where $\mu_0$= Ion Mobility, $V$= Voltage, P = Working pressure, $P_0$ = Atmospheric Pressure, D = gap (for plane electrodes) or $D^2 = [(a^2 - b^2)/2] \ln (a/b)$ (for cylindrical geometry with *a, b* = radius of the inner and outer electrode).

The ion mobility $\mu_{ion}$ is for typical gases $\mu_{ion} = 1.0 – 1.5$ cm$^2$ atm/(V·s) which gives for $D = 1$ cm, $P = 1$ atm, V = 1.5 kV a response time of $t_\mu \approx 0.4$ - 1 ms, which is often too slow for BLMs. The electron mobility $\mu_e$ is about $10^3$ times faster than for ions, therefore

$$t_e \approx 0.4 - 1 \; \mu s.$$

The signal of an ionization chamber is a combination of both and is shown in Fig. 15. Further improvement in speed was achieved for the LHC type of ionization chamber. Here pane electrodes have a smaller gap in a meander constellation to fill a volume of 1.5 ltr. A response time of 0.3 μs is achieved [44], [45]. A comparison between the SNS and LHC chambers is shown in Fig. 16.

Another important aspect is the dynamic range of the ionization chamber. The upper limit is given by the nonlinearity due to the recombination rate at high dose; the typical chamber current in such a case is a few hundred μA. The lower limit is given by the dark current between the two electrodes. A very careful design of the chamber is necessary to allow very low dark currents in the order of tens of pA or better. This gives a dynamic range of $10^6 – 10^8$. Note that such a high dynamic range needs some special signal processing. Solutions like logarithmic amplifiers, high ADC resolution, current to frequency conversion and counting schemes [45] are reported.

Ionization chambers can be built from radiation-hard materials like ceramic, glass and metal with no radiation or time ageing. Special care has to be taken for the feedthroughs. Up to more than $10^8$ rad can be tolerated by a careful design focused on radiation hardness. Air-filled ionization chambers need nearly no maintenance; a leakage in $N_2$ filled chambers is not critical. Air-filled chambers give very little reason for maintenance. However, it might become necessary to periodically verify the connection to each corresponding channels of the electronic system and the signal quality of all detectors by a radioactive source [46].

[1] Avoid electronegative gases ($O_2$, $H_2O$, $CO_2$, $SF_6$ …), they capture electrons before reaching the electrode. Nobel gases have negative electron affinities (Ar, He, Ne), better for the proportional region.

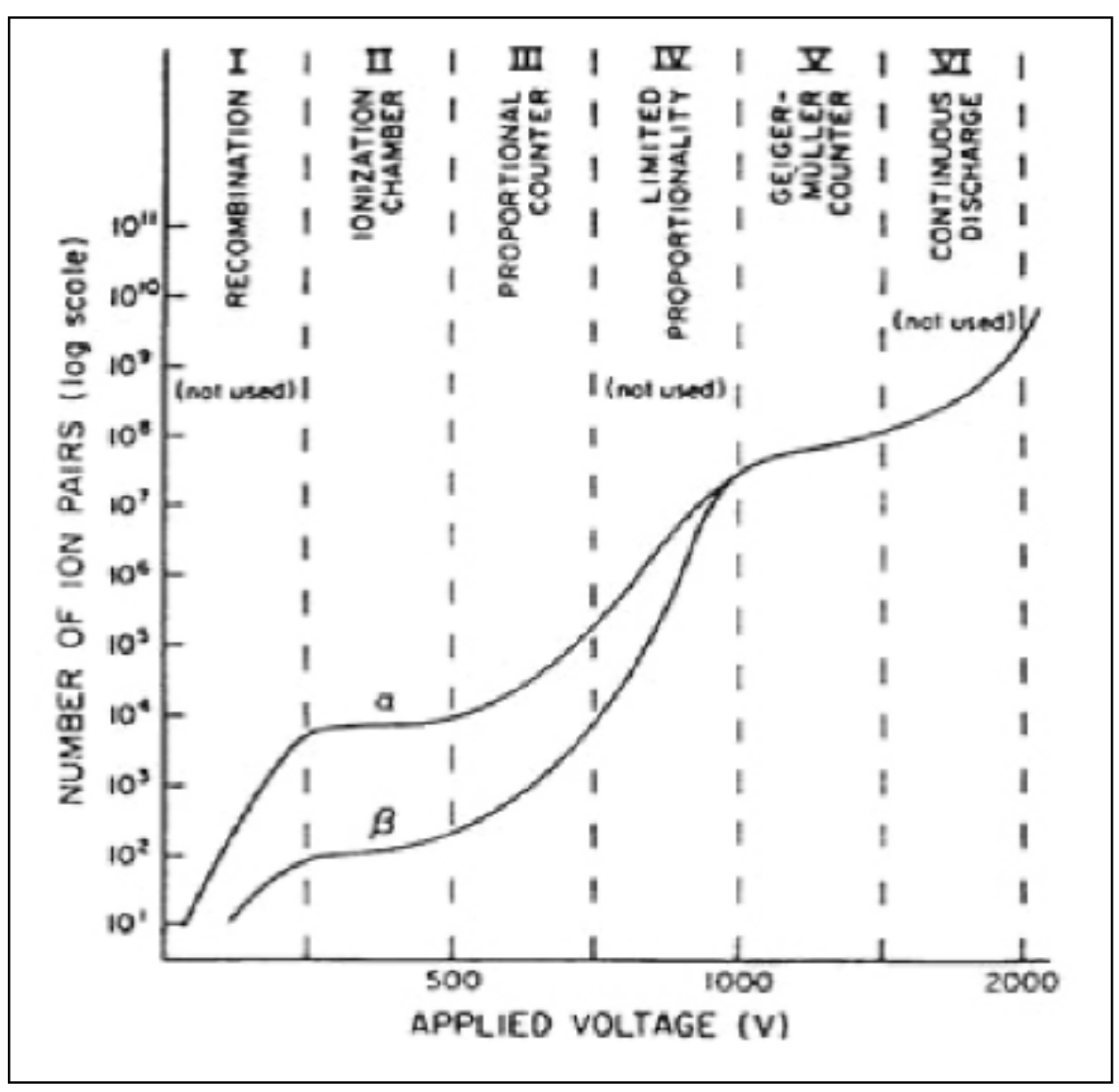

**Fig. 14:** Regions of gas-filled ionization chambers for α and β particles, from [38].

### *4.1.1 Calibration and sensitivity of ionization chambers*

Note that for comparison and simplicity the sensitivity $S_{BLM}$ [C/rad] of various types of BLMs is calculated for 1 rad dose created by MIPs in the following chapters.

The number of electrons (ne) produced in the gap D of an ionization chamber by one minimum ionizing particle (MIP) is:

$$n_e = \frac{D \cdot \rho}{W} \cdot \frac{dE}{dx}$$

with dE/dx from Eq. (1) (Bethe Bloch)[2], ρ = density of the medium and W = W-value = energy to create one electron/ion pair [47]. ρ and W can be collected from textbooks. Note that W is consider-ably higher than the usual “first ionization potential” given in most textbook tables. W is about constant for many gases and radiations.

Example: Argon filled chamber:

$\rho = 1.661 \cdot 10^{-3}$ g/cm$^3$ (20 $^0$C, 1 atm), dE/dx = 1.52 MeV/(g/cm$^2$)

=> $n_e \approx 100$/cm · D [e/MIP].

The intrinsic sensitivity $S_{BLM} = S_{ION}$ (normalized to 1 rad) of an ionization chamber depends on its size. Let’s assume as an example a 1 ltr. Argon filled chamber with 100% charge sampling efficiency:

$$S_{Ion} = 100\frac{erg}{g} \cdot \frac{1eV}{1.6 \cdot 10^{-12}erg} \cdot \frac{1e^-}{26eV} \cdot \frac{1.66 \cdot 10^{-3}g}{cm^3} \cdot 1000\,cm^3 \cdot \frac{1.6 \cdot 10^{-19}C}{e^-} = 638\frac{nC}{rad}$$

| 1 rad |·| eV → erg |·| W |·| ρ |·| 1 ltr. |·| $e^-$ charge |

[2] Cross section for nuclear interaction is about 5 ·$10^{-6}$ times the ionization cross section ($10^{-16}$ cm$^2$). Cross section for excitation is about $10^{-17}$cm$^2$. Both do not contribute significantly to the energy deposition. Rutherford (nuclear) scattering does not produce significant energy transfer, but angular spread.

This value correlates well with the measured values in the table of Fig. 16. A beneficial characteristic of ion chambers is that their sensitivity is determined by geometry (size), and that the sensitivity is relatively independent on the applied voltage.

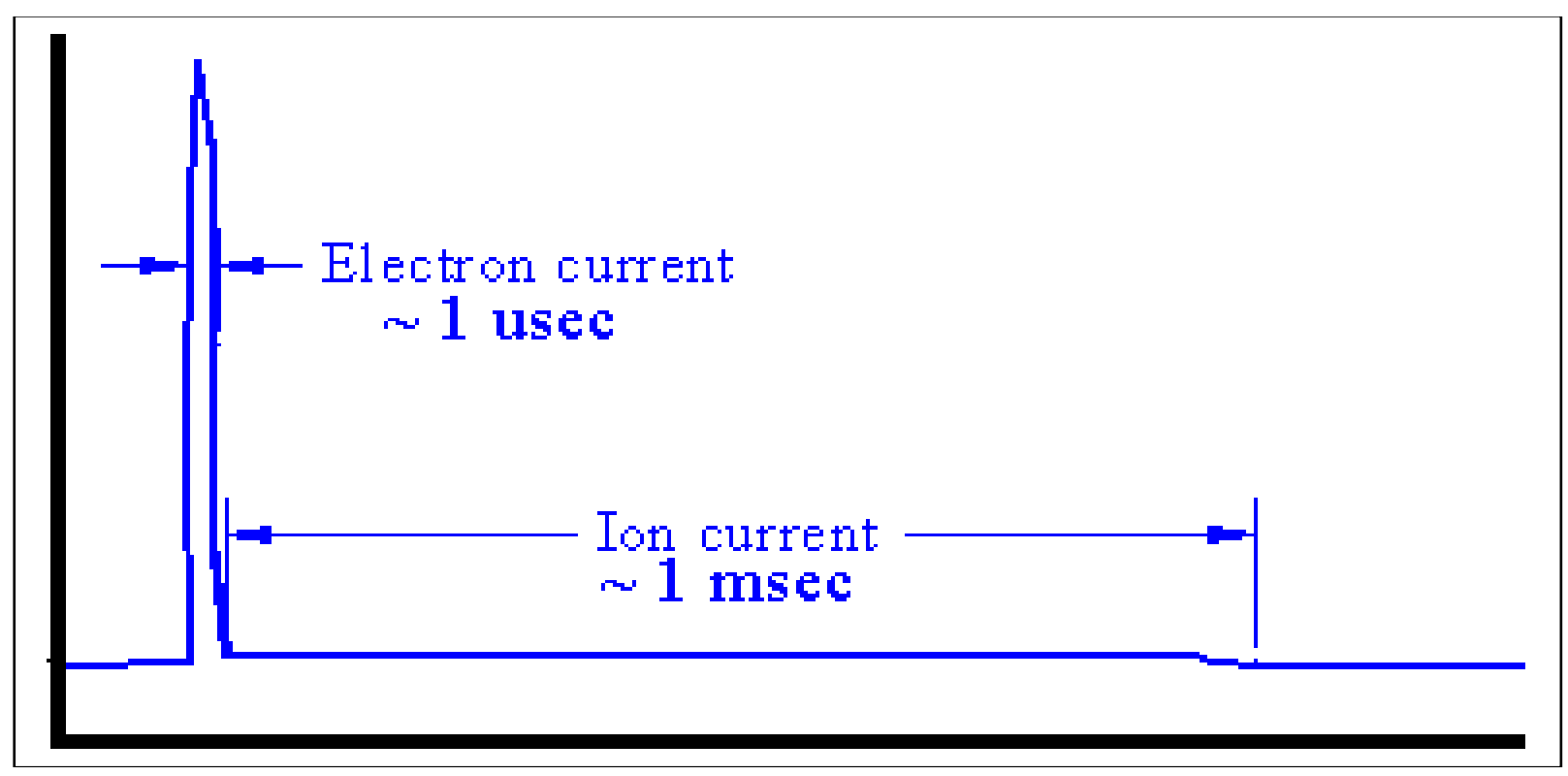

**Fig. 15:** Typical signal of an ionization chamber, from Ref. [43]

| LHC | TEVATRON/RHIC/SNS |
|---|---|
| T = 0.3 μs ($t_{fall}$ 200μs) | T = 10μs => 3 μs ($t_{fall}$ 560 => 72 μs) |
| 1.5 ltr $N_2$ at 1.1 bar | 0.11 ltr Ar at 1 bar |
| V = 800 – 1800 V | 500 - 3500 V |
| **Dynamic range >$10^8$ (>$10^{-12}$ – <$10^{-3}$ A)** | Dynamic range >$10^6$ 300 pA – 500 μA |
| Leak current <1 pA | Leak current 10 pA => <100 fA |
| S: 156 pA/(rad/h) ($Cs^{137}$) (560 nC/rad) | 19.6 pA/(rad/h) ($Cs^{137}$) (70 nC/rad) |
| Collection effeciency: >90 % | Collection effeciency: 77% -> 92 % |

**Fig. 16:** Comparison between LHC and SNS ionization chamber (courtesy of B. Dehning and M. Stockner, CERN). For the SNS chamber some improvements from the original Tevatron and RHIC chamber [48], [49] are indicated by '=>'.

### 4.2 Long ionization chamber

A 'short' ionization chamber described in Section 4.1 covers only a small part of an accelerator; therefore, a large number need to be installed to get good coverage. To overcome this problem in linacs or transport lines, a long and gas-filled coaxial cable (e.g. Andrew® HJ4.5-50 HELIAX) can be used as an ionization chamber. In 1963, Panowsky [50] proposed a BLM system for SLAC consisting of one long (3.5 km) hollow coaxial cable. It was an industrial RG-319/U cable with a diameter of 4.1 cm, filled with Ar (95%) + CO2 (5%). It was named Panowski's long ionization chamber or PLIC. It was mounted on the ceiling along the linac, about 2 m from the beam. This BLM system worked for more than 20 years and was upgraded for the SLC [51]. Nearly the entire SLC is covered by a few PLICs.

Position sensitivity is achieved by reading out at one end the time delay Δt between the direct pulse and the reflected pulse from the other end (see Fig. 17). The time resolution due to the electron collection time of $t_e \approx 270$ ns and the dispersion in the cable is about 50 ns (~ 15 m) for 6 km long cables, for shorter PLICs about 5 ns (~ 1.5 m) was achieved. This principle of space resolution works for one-shot (or one-turn) accelerators (and transport lines) with a bunch train much shorter than the length of the cable and with relativistic particles. For particles travelling significantly slower than the signal in the cable (≈ 0.92c), the resolution of multiple hits in the cable becomes difficult. In this case and for circular machines it is necessary to split the cable. Each segment has to be read out separately, with spatial resolution approximately equal to the length of the unit (see Fig. 18). This was done at various linacs, e.g., ELBE [52], [53] and in the BNL 200 MeV linac [54], but also in some rings, e.g., in the AGS ring, Booster and transport lines [55], [56]. In the KEK-PS, 56 air-filled cables with a length of about 6 m have been installed. Using amplifiers with a variable gain, a dynamic range of 104 is achieved [57].

Long ionization chambers made of commercial cables are simple to use, cheap, and have a uniform sensitivity. The isolation is not very radiation hard; nevertheless, these cables were used in SLAC for more than 20 years without serious problems.

#### *4.2.1 Calibration and sensitivity of long ionization chambers*

The sensitivity $S_{long_ion}$ of a long ionization chamber depends on its geometry and diameter. A sensitivity of

$$S_{long_ion} \approx 200 \text{ nC/rad/m}$$

can be expected [49] for a 7/8 inch Argon filled (1 atm) cable. Commercial cables have a typical resistance of $R=10^{13}$ Ωm between the inner and outer conductor, with a bias of U=100 V this gives a dark current of $I_{dark} = U/R = 10$ pA/m. For long cables in the order of km, this limits the dynamic range of such a BLM. Reference [58] proposed a cable with guard electrodes to improve the dark current and therefore its dynamic range.

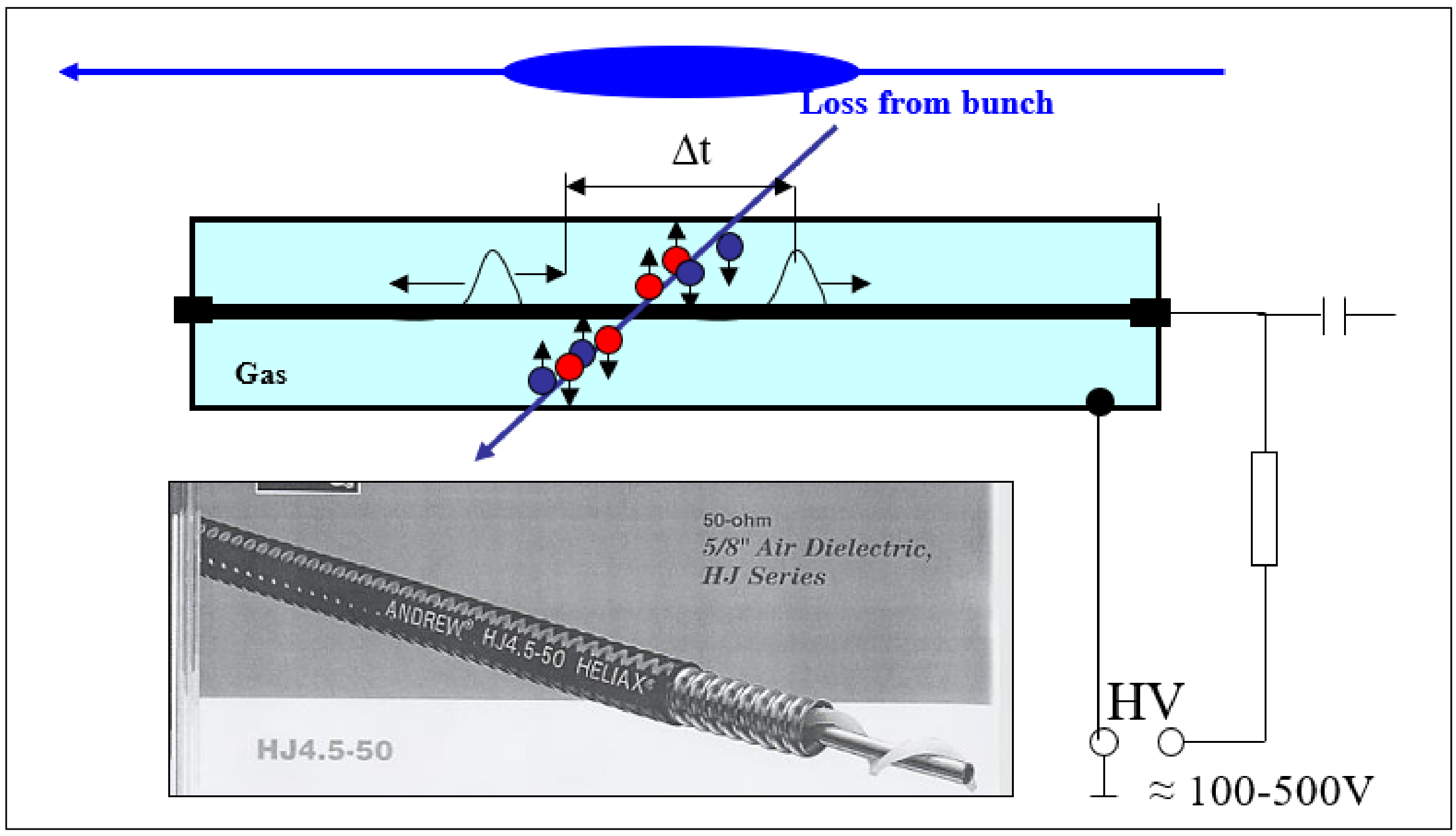

**Fig. 17:** Principle of position measurement with long ionization chambers. The inset picture is an example of a commercial HELIAX cable.

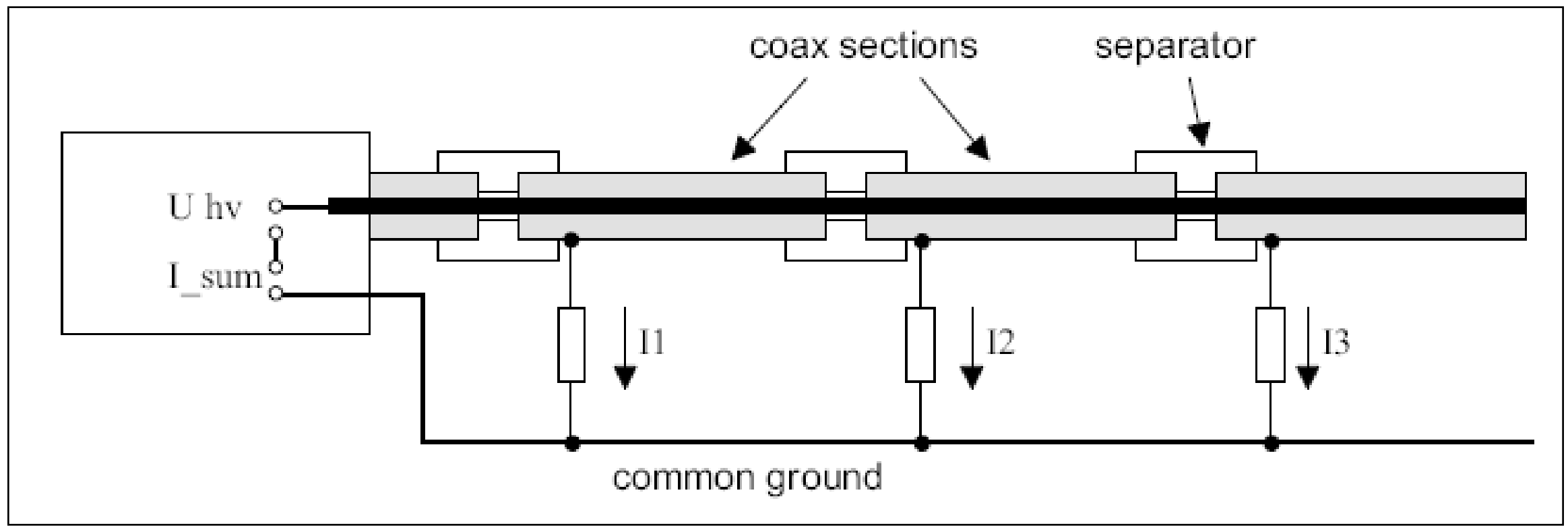

**Fig. 18:** Principle of a sectioned cable [52]. The signal can be measured at I1, I2, etc.

### 4.3 Solid-state ion chamber – PIN photodiode and CVD Diamonds

The required energy to create an electron-hole pair in a semiconductor is about 10 times smaller than creating an electron/ion pair in gas. In addition, the density of a semiconductor is about three orders of magnitude larger than for gas (at 1 atm). Therefore, a MIP creates in a semiconductor much more signal per path-length than in a gas volume. Let us assume as an example a large-area silicon-PIN photodiode used as a 'solid-state ionization chamber' with an active area A = 1 $cm^2$. Such diodes are commercially available from Hamamatsu (http://www.hamamatsu.com), for example. A MIP creates free charges in the depletion (or 'intrinsic' or I-) layer between the high doped p+- (P-) layer and low doped n- (N-) layer of the P-I-N diode (see Fig. 19). The width $d$ of the depletion layer is given by

$$d_{n,p} \approx 0.5 \cdot \sqrt{U[V] \cdot \rho_{n,p}[\Omega m]},$$

where $\rho_{n,p}$ is the specific resistance of the n- or p- doped Silicon and U is the applied bias voltage. Since the low doped n-layer has a much higher resistance ($\approx$ 5-10·$10^3$ Ωm) most of the depletion layer is located in the n-layer. Assuming a reversed bias voltage of 25 V the depletion layer has a width of about 100 – 200 μm. At about 30-40 V the width d saturates due to the limited thickness of the Si-layers. The energy deposition in Si for a MIP is d$E$/d$x$ = 3.7 MeV/cm/MIP. The energy to create an electron e (-hole pair) in Silicon $E_e$ is only 3.6 eV/$e^-$. Therefore $N_e$= d$E$/d$x$ · d/$E_e$ = 1.03 ·$10^4$ e/MIP are created in a 100 μm depletion layer. The electrons are guided to the n+-layer with an efficiency of >> 80%. The electron mobility $\mu_e$ is higher than for gas, typical electron mobility for doped Silicon at room temperature (300 K) is

$$\mu_e \approx 0.3 \text{ m}^2/(\text{V}\cdot\text{s})$$

(http://www.ioffe.ru/SVA/NSM/Semicond/Si/electric.html); (for holes: $\mu_{hole} \approx 450$ $cm^2$/(Vs)). Owing to the small gap $d$ = 200 μm and a high electric field $E$ = 25 V/200 μm = 1.25 ·$10^5$ V/m the transit time $t_e$ of the signal is very fast:

$$t_e = d/(\mu_e E) \approx 5.3 \text{ ns}$$

The capacity $C$ of the diode is proportional to $A/d$ and to 1/√V and saturates with the width of the depletion layer at around 30 V at values between $C \approx$ 10 – 100 pF, depending on the area A of the diode. Therefore, the rise time of the signal defined by the time constant τ = C · $R_L$ ($R_L$ = load resistance,) might be in the same order of magnitude: Assuming $R_L$ = 50 Ω:

$$\tau = 100 \text{ pF} \cdot 50\Omega = 5 \text{ ns}.$$

Owing to the finite resistance between the two electrodes ($p^+$ and $n^+$) of the diode a dark current of a few nA is typical. The dark current is proportional to the applied bias voltage and does not depend on the thickness of the depletion layer. However, radiation damage (up to $10^6$ rad [59]) leads to an increase of the dark current while most of the other parameters keep unchanged.

Note that also not too strong magnetic fields do not influence the charge collection in PIN-Diodes. Therefore, they can work as BLMs in stray fields of magnets.

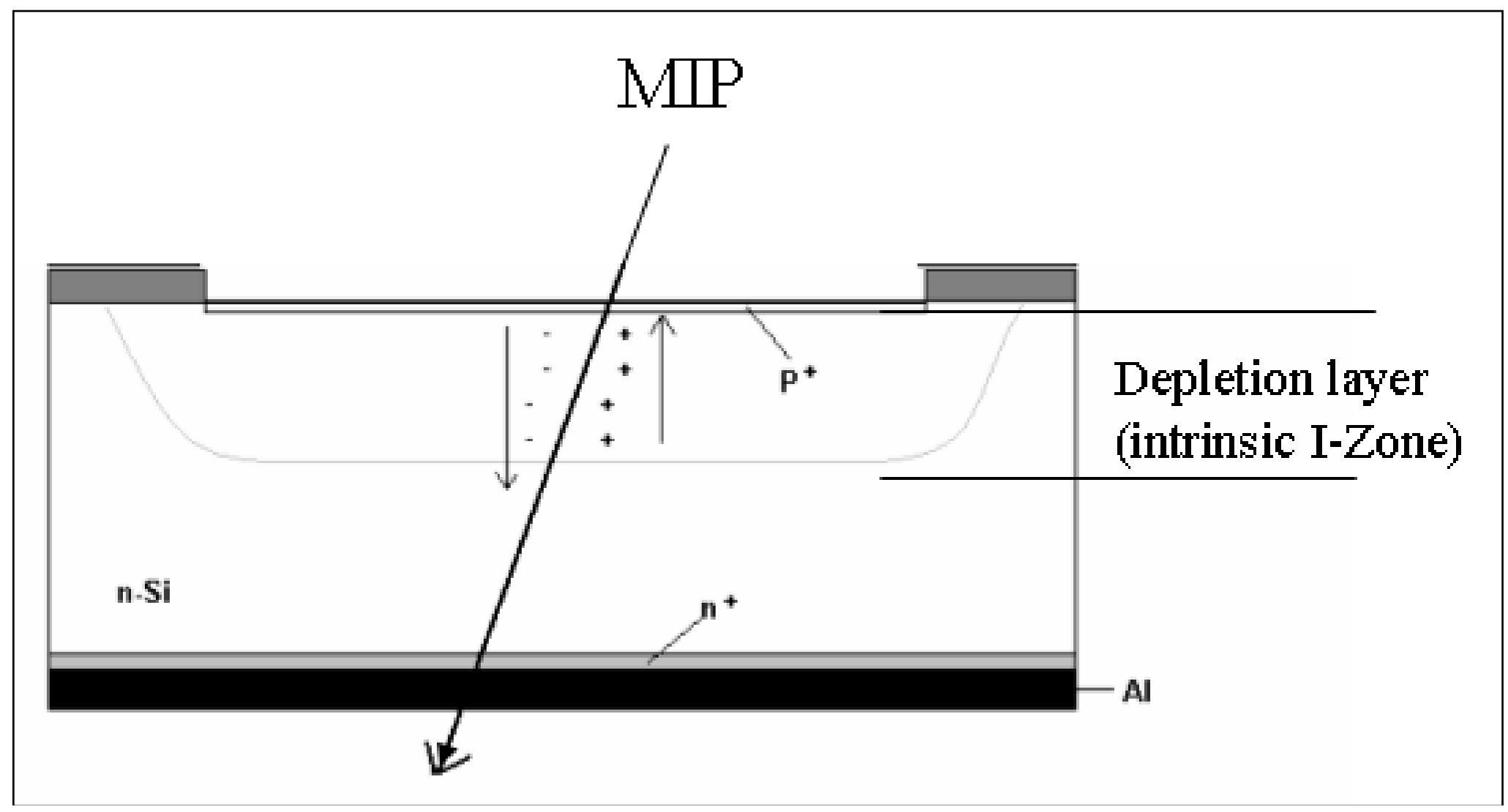

**Fig. 19:** Schematic of the signal creation in a PIN diode by a MIP

### *4.3.1 Calibration and sensitivity*

The sensitivity $S_{PIN}$ can be calculated with the help of Eq. (2); 1 rad = $3.1 \cdot 10^7$ MIP/cm$^2$. Assuming a diode with $A$ = 1 cm$^2$ and a depletion layer width $d$ = 200 µm, resulting in $N_e = 2 \cdot 10^4$ e/MIP (assuming nearly 100% collection efficiency), we get:

$$S_{PIN} = 2 \cdot 10^4 \frac{e}{MIP} \cdot 3.1 \cdot 10^7 \frac{MIP}{cm^2 \cdot rad} \cdot 1.6 \cdot 10^{-19} \frac{C}{e} = 100 \frac{nC}{rad \cdot cm^2}$$

Operating such a diode in a current mode like an ionization chamber the sensitivity is

$$S_{PIN} = 100 \text{ nC/rad}$$

for a 1 cm$^2$ Diode with a depletion layer width of $d$ = 200 µm.

Examples of a current mode PIN photodiode BLM system can be found in [60].

### *4.3.2 Pin diode in counting mode*

Circular electron accelerators emit hard synchrotron radiation (SR). The radiation interacts with the BLMs and a separation between SR background and the beam loss distributions using BLMs in current mode technique is difficult to achieve [60]. HERA, for example, was an accelerator with an electron and a proton ring in the same tunnel, operating at the same time. Protection of the superconducting proton magnets from beam-loss-induced quenches must rely on a BLM system which sees only the proton beam losses and not the SR background. The (hadronic) shower created by beam losses includes a large number of charged particles, in contrast to the photons of the SR. The HERA BLM system consists of two PIN photodiodes mounted close together (face to face) and read out in coincidence [27]. Thus, charged particles crossing through the diodes give a coincidence signal, while photons which interact in only one diode do not. In contrast to the charge detection of most other BLM systems, coincidences are counted while the count rate is proportional to the loss rate as long as the number of overlapping coincidences is small. Counting of MIPs crossing both diodes have a few implications:

1) Both channels need a discriminator to suppress dark counts due to noise. Since the signal of one MIP is still weak, the threshold cuts also some of the MIP signals. The efficiency of such a BLM for MIPs was measured by [61], [62]. It was found to be about $\varepsilon_{count}$ = 30-35 %/MIP for the coincident readout of the BLM which includes also the readout electronic. Assuming a dose-rate of 1 rad/s = $3.1 \cdot 10^7$ MIP/s/cm$^2$ Eq. (2) gives a sensitivity of

$$S_{count} = \varepsilon_{count} \cdot 31\ \text{MHz/cm}^2\text{/rad} = 9.3\ \text{MHz/rad}$$

for a diode with $A$ = 1 cm$^2$.

2) The noise (dark count rate) is very small, typically < 0.01 Hz due to the coincident readout.
3) The detector cannot distinguish between one or more MIPs crossing both diodes at the same time. The shortest signal length is defined by the response time of the diodes, but in practice it is defined by the readout electronics. An efficient counting type of BLM should have a signal length shorter than the bunch distance, so that the maximum measured loss rate is the bunch repetition rate of the accelerator (at HERA 10.4 MHz). Saturation effects occur even before the maximum rate, but they can be corrected by applying Poisson statistics [63].
4) The dynamic range lies between the dark count rate of < 0.01 Hz and the maximum rate (e.g. HERA: 10.4 MHz) and might reach $10^9$. However, remaining background counts from synchrotron radiation have to be taken into account:

   Two effects contribute to the background rate:

   a. A high energy photo- or Compton- electron produced by a SR photon conversion in the first diode can reach the other diode and creates a coincident signal in the two diodes. One can find in [64], [65] a detailed description how a copper inlay between the two diodes reduces the probability of background counts due to photo- or Compton-electrons. The efficiency of the BLM to beam losses is not changed by this inlay.
   b. The high photon rate of SR gives a higher probability for the coincident conversion of two photons in the two diodes. An efficiency of the BLM of $\varepsilon_\gamma = 3.5 \cdot 10^{-5}$ was measured [66], [67] which is already 4 orders of magnitude below the MIP efficiency. A reduction of the high photon flux by a lead hat helps in 2 ways [67]: First the probability for a coincident conversion of two photons is additionally reduced and secondly the background rate of Photo- or Compton electrons is reduced.

The PIN diode BLM system (Fig. 20) at HERA was successfully operated between 1992 and 2007 without significant problems or radiation damage.

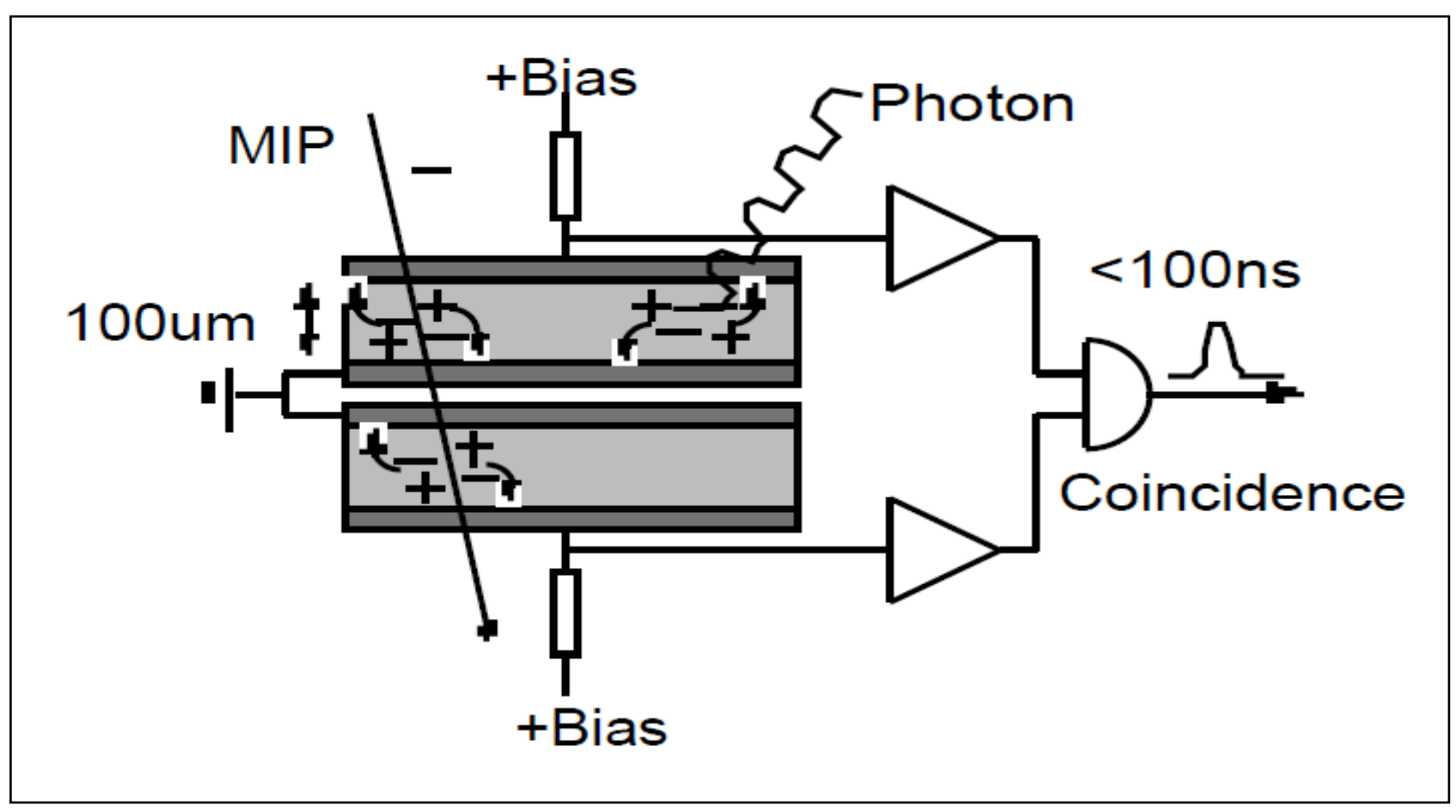

**Fig. 20:** Operating principle of counting mode PIN diode BLM, from Ref. [68]

#### *4.3.3 CVD Diamonds Sensitivity*

CVD[3] Diamonds offer a superior robustness and radiation resistance. The signal speed is almost as fast as for PIN Diodes. Even in cryogenic environment the signal length is about 3.6 ns [69], that is in many cases fast enough to separate losses from individual bunches. The sensitivity $S_{CVD}$ can be calculated with the help of [70] and [71], showing that the mean charge created by a MIP in CVD Diamond is:

$$S_{CVD} = 0.44 * S_{PIN} = 44 \text{ nC/rad}$$

for a 1 $cm^2$ Diamond with a width of $d$ = 200 μm.

More example of Diamond based detection systems can be found in Refs. [72], [73].

### 4.4 Secondary emission monitors

Ionization chambers have the disadvantage of being slower than the bunch distance in most accelerators (except for PIN diodes in current mode). Counting mode devices have to integrate the counts over a lot of bunches to get a statistically relevant signal. A simple, robust and cheap BLM is a secondary emission chamber. As described in Section 3.2, secondary electrons are emitted from a surface due to the impact of charged particles. An efficiency of a few per cent was measured for many metal surfaces (Fig. 12). That means that 100 MIPs crossing an area will produce only a few electrons. Since Secondary Electron Emission (SEM) is a very fast effect, such a monitor is very fast (≈ ns). But its very low sensitivity makes it useable only in high radiation fields; with the additional advantage that it consists of nothing more than a few layers of metal (see Fig. 21). Therefore, it is a very radiation hard monitor. The monitor has to be evacuated to avoid contamination of the signal due to gas ionization! Since the efficiency of gas ionization is much higher, a gas pressure of better than $10^{-4}$ mbar should be achieved to get < 1% signal from ionization. Especially in high radiation fields, gas ionization will lead to nonlinearities while SEM is a very linear process over a wide range of intensities [22], [26]. Unavoidable ionization outside the detector at the feedthroughs and connectors limits the linearity at the lower end of the signal [45] .

A SEM multiplier extends the use of SEM BLMs to small radiation intensities. As far back as 1971 CERN [74] used Aluminium Cathode Electron Multipliers (ACEM) for beam loss measurements. This device is a photomultiplier tube where the photocathode is replaced by a simple aluminium cathode. The SEM electrons are guided to dynodes where they are amplified; amplifications up to 106 were possible. An example for recent use of ACEMs can be found at FLASH [75].

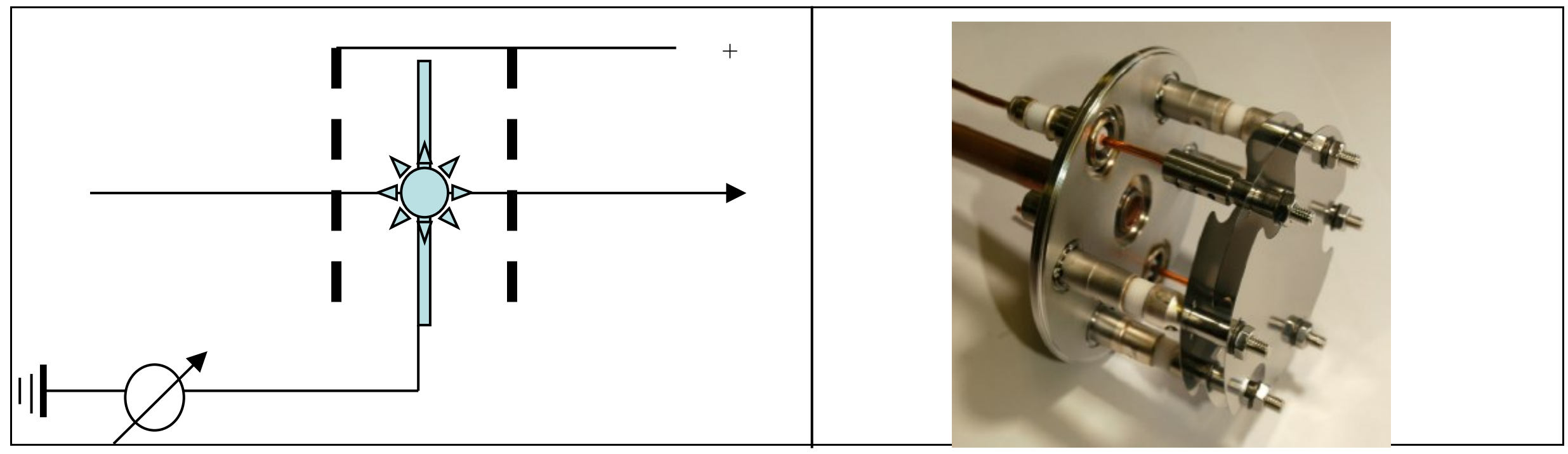

**Fig. 21:** Left: a sketch of a SEM monitor. The SEM foil in the middle is surrounded by positive charged grids to avoid the repulsion of the emitted SEM electrons. Right: LHC SEM BLM with gas sealed cover removed (from Ref. [45]). A NEG ST707 foil keeps the gas pressure below $10^{-4}$ mbar.

[3] Synthetic Diamond grown by Chemical Vapor Deposition (CVD) method

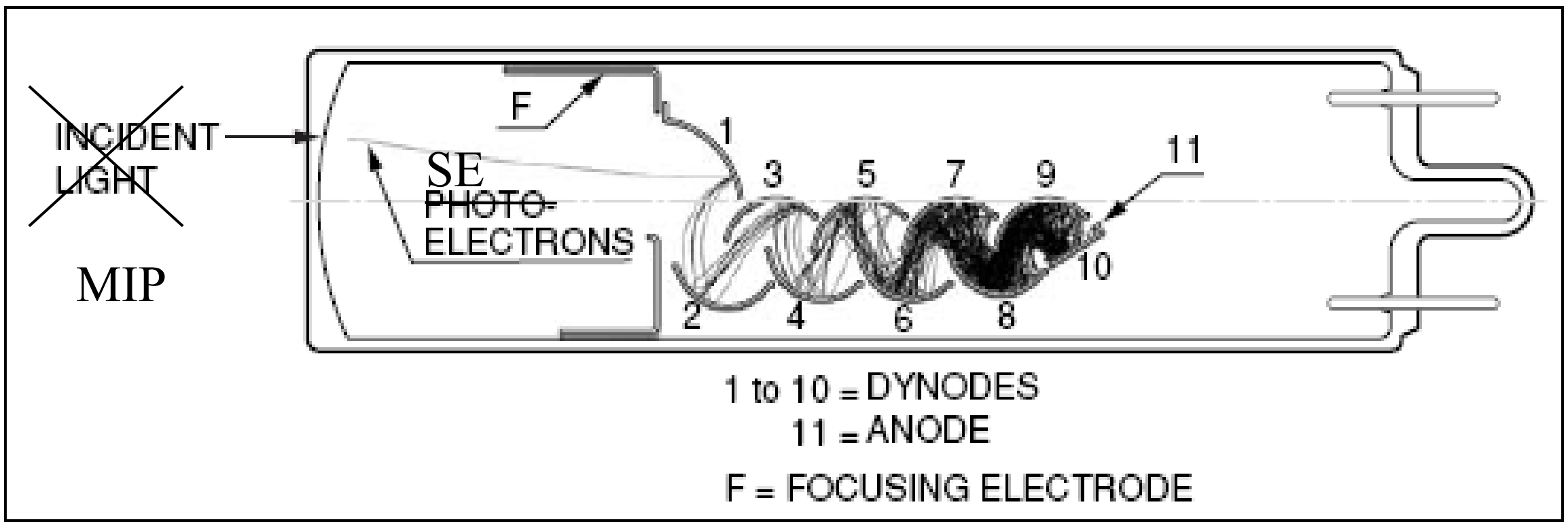

**Fig. 22:** Sketch of an aluminium electron multiplier [76]

#### *4.4.1 Calibration and sensitivity*

SEM chambers can be built in any size. Let's assume a SEM yield of $\mu_{SEM}$ = 5 % = 0.05 e/MIP (from Ref. [26]). With Eq. (2) this gives:

$$S_{SEM} = 3.1 \cdot 10^7 \frac{MIPs}{cm^2\, rad} \cdot 0.05 \frac{e}{MIP} \cdot 1.6 \cdot 10\text{-}19 \frac{\mathrm{C}}{\mathrm{e}} = 0.25 \frac{pC}{rad\, cm^2}$$

The gain of the ACEM tube can be adjusted by the applied high voltage and varied by a few orders of magnitude. This has to be taken into account. Let's take as an example the diameter of the ACEM Type 8941, which has an area of $A$ = 8 cm$^2$. Therefore, the sensitivity $S$ for this BLM can be written as:

$$S_{ACEM} = 2 \frac{pC}{rad} \cdot \left( ACEM_{gain} \right)$$

Since the gain may drift as with a photomultiplier tube, care has to be taken to keep the high voltage and the gain stable. Also, radiation damage in ACEMs may lead to a reduction of the gain (not reported up to now).

### 4.5 Scintillation detectors

SEM-based BLMs are very fast but still have a moderate sensitivity. An equivalent speed but much higher sensitivity can be expected from scintillation counters. This is a combination of a scintillating material and a photomultiplier tube (PMT) (or a SiPMT to overcome gain drift, false SEM signals, magnetic fields, costs [77]). Large-area plastic (organic) and liquid scintillators are available. Plastic scintillators, in particular, can be modulated in nearly any kind of shape and size while inorganic scintillators are expensive and limited in size. Descriptions of how the scintillation process works can be found in Ref. [38] and in various text books, e.g., Refs. [40], [41]. One important fact should be stressed here: *Note that the flux density of photons from the scintillator into the light guide is 'incompressible'! The cross section of the scintillator should not be larger than the cross section of the light guide.* However, large scintillators can be useful to enhance the solid angle of beam loss detection if the resulting radiation is not uniformly distributed. This is often true if the BLM is located very close to the beam pipe where the radiation is peaked into a solid angle. Examples are shown in Fig. 23. Typically, a thin layer of scintillator (1–3 cm) is sufficient to ensure sensitive loss detection.

The light L produced by a charged particle in a scintillator is for small dE/dx (e.g. for MIPs): dL/dx = $R_s$ dE/dx, where $R_s$ is the ratio of the average number of emitted photons to the energy of the incident radiation absorbed by the scintillator. $R_s$ is tabled for typical scintillator materials by various commercial suppliers, e.g. [78]. For a typical plastic scintillator, e.g. NE102A, one needs energy to create one photon of $1/R_s \approx 100$ eV/photon. The scintillation light is then transported from the scintillator layers to the photosensitive device via light guides. Attenuation occurs due to internal absorption and mainly due to missing internal multiple reflections. The light is reflected at the surface of the scintillator and the light guide due to their higher refraction coefficient $n_1$. Once the angle of the light relative to the surface is larger than the the critical angle given by $\Theta_c = \arcsin(n_2/n_1)$ ($n_2/n_1 \leq 1$, $n_2$ = refractive index of the less dense medium, here air $n_2$=1) the light leaves the material. A reflective foil, e.g. aluminum foil or with paper, reflects the light back, but with some higher absorption. Note that the main process of the light transport is still the internal total reflection defined by the higher refraction index. Any distortion will lower the light transport efficiency. A good matching between the light guide and the PMT is also very useful. Therefore:

1) Match the size of the light guide and the photocathode.
2) Optical grease between light guide and PMT helps to couple out the light.

A carefully designed scintillator assembly has an overall collection efficiency of about:

$$\varepsilon_{coll} \approx 60\%.$$

3) The emission spectrum of the scintillator and the spectral response of the PMT should match.

Well matched spectra result in a photon-to-photoelectron conversion efficiency of the photocathode of $\varepsilon_{cath} \approx 30\% = 0.3$ e/photon. These photoelectrons are amplified in the PMT with a gain between $10^5 \leq PMT_{gain} \leq 10^8$, depending on the type of PMT and the applied high voltage. The shape of the scintillation pulse is characterized by a fast rise time of the order of 1 ns and a decay time of a few ns [79].

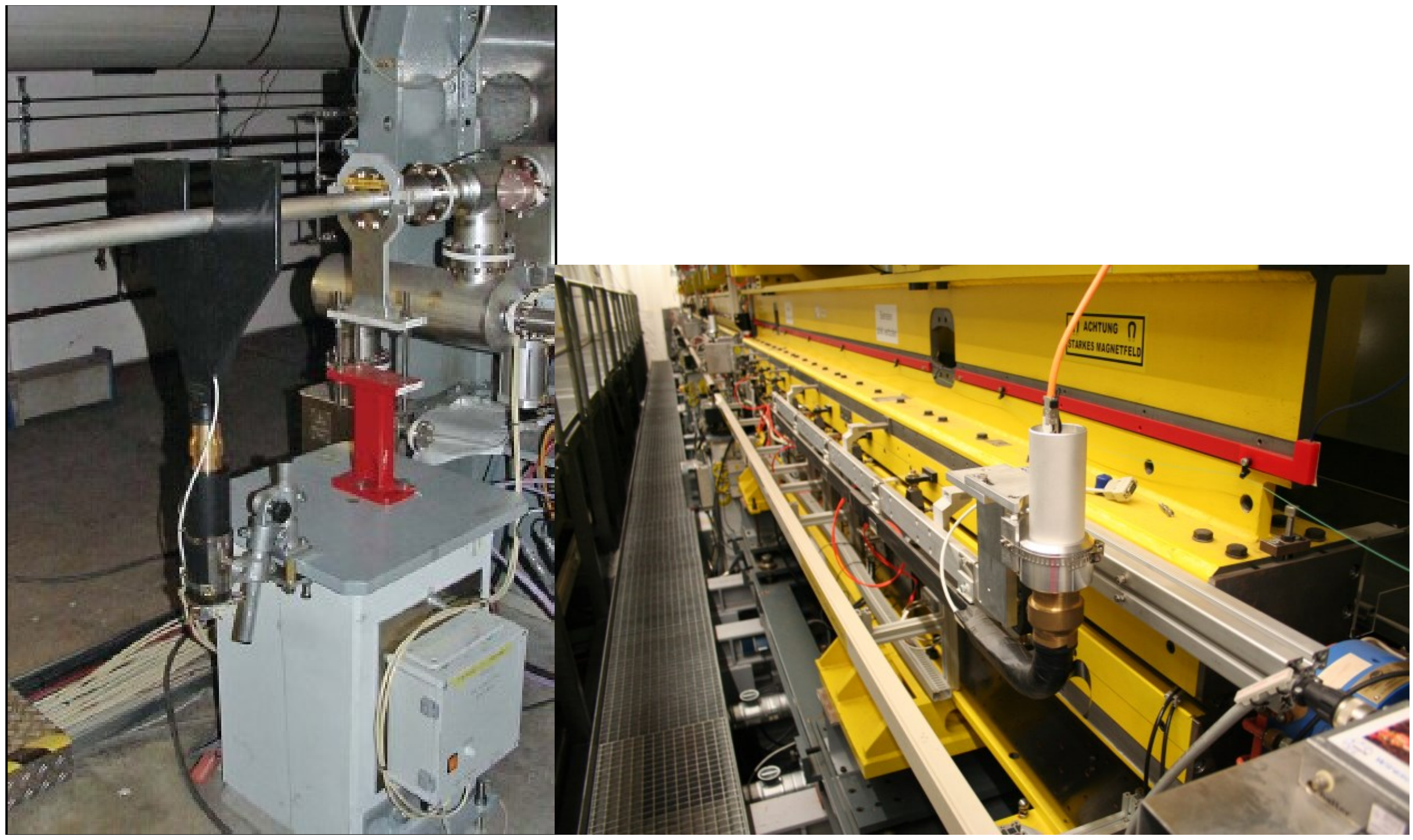

**Fig. 23:** Scintillator BLMs at FLASH (DESY); Left: Surrounding the beam pipe; Right: Long scintillator along an undulator. Courtesy L. Fröhlich, T. Wamsat, DESY.

#### 4.5.1 Calibration and sensitivity of scintillation counters

Since the sensitivity depends on the type and size of the scintillator, let's assume NE102A as material. NE102 has a density ρ = 1.032 g/cm$^3$ (Weight = 1032 g/ltr.) and a light output of $R_s$ = 0.01 photon/eV. A deposition of 1 rad = 100 erg/g into this volume creates N number of photons:

$$N = \frac{100\,erg}{g \cdot rad} \cdot 1032 \frac{g}{ltr.} \cdot \frac{1\,eV}{1.6 \cdot 10^{-12}\,erg} \cdot R_s \approx 6.5 \cdot 10^{14} \frac{photons}{rad\,ltr.}$$

Therefore, the sensitivity $S_{scint}$ for Ne102A scintillator is:

$$S_{Scint} = 6.5 \cdot 10^{14} \frac{photons}{rad} \cdot \varepsilon_{coll} \cdot \varepsilon_{cath} \left[ \frac{elctrons}{photon} \right] \cdot \frac{1.6 \cdot 10^{-19}\,C}{electron} \approx 18 \frac{\mu C}{rad\,ltr.}$$

Assuming a counter with a volume of V=1 ltr. = 1000 cm$^3$ (the dimension might be 25·25·2 cm$^3$) connected to a PMT this gives a sensitivity $S_{scint}$:

$$S_{Scint} = 18 \frac{\mu C}{rad} \cdot PMT_{gain}$$

Note that the light transmission through the scintillator (and the Plexiglas light guide, ecoll) changes on account of radiation damage. This depends strongly on the scintillator material, but for organic scintillators a typical value can be assumed: the transmission decrease to 1/e of its original value happens after about 0.01–1 MGy (1–100 Mrad) collected dose. Liquid scintillators are somewhat radiation harder and have about the same sensitivity. An older example of liquid scintillator BLMs can be found in Ref. [80]. Complete liquid scintillator detectors are also commercially available [81]. Note that inorganic scintillators like BGO or CsJ(Tl) have about a factor 10–50 higher sensitivity due to their higher density (factor ≈ 5) and larger $R_s$ (factor ≈ 5). But their radiation hardness is poor and large-sized crystals are very expensive.

The gain of the same type of photomultipliers (PMTgain) varies within a factor of 10. Therefore, a careful inter-calibration of the BLM sensitivities is necessary by adjusting the high voltage (HV). The drift of the gain is a well-known behaviour of PMTs. A stabilized HV source and continuous monitoring of the photomultiplier gain over the run period are necessary to keep the calibration error small.

### 4.6 Cherenkov light

Scintillation detectors are sensitive to charged particles *and* to X-rays. A Cherenkov detector is nearly insensitive to X-rays because the Cherenkov effect occurs only when the velocity of a charged particle traversing a dielectric medium is faster than the speed of light in that medium. In that case photons are emitted at an angle with respect to the trajectory of the particle defined by the velocity β of the particle and the refraction index n of the medium. The light can be focused on a PMT to build a BLM (see Fig. 24). The Cherenkov material (e.g., synthetic fused silica) has good radiation hardness up to about 1 MGy [82]. An example for a Cherenkov based BLM system can be found in Ref. [83].

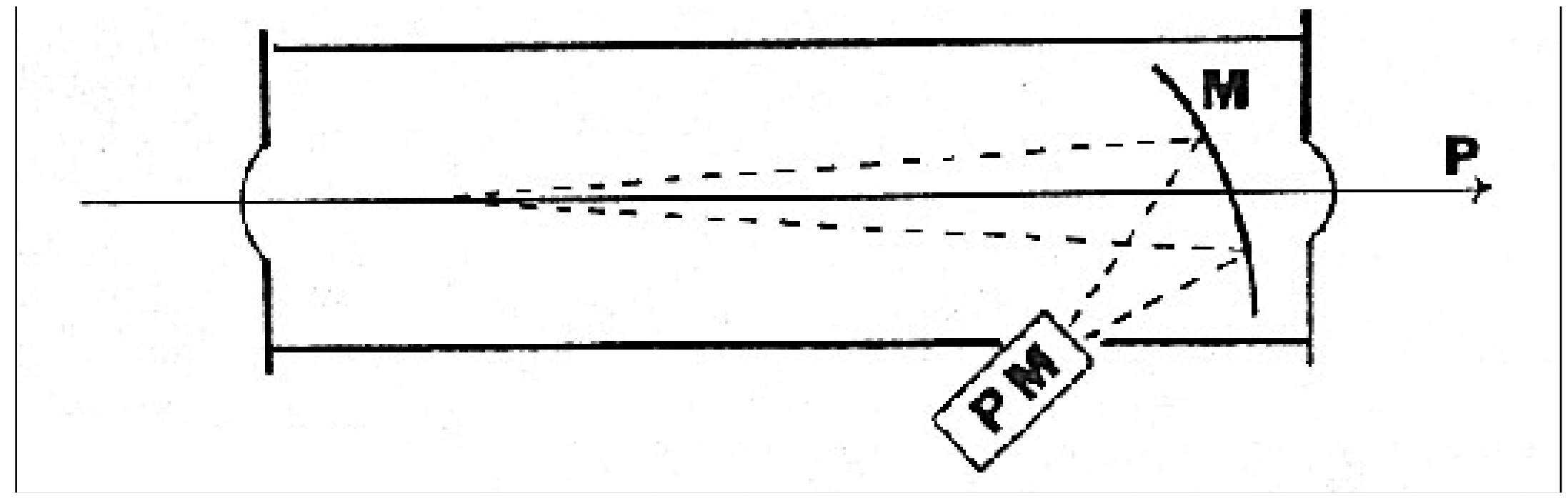

**Fig 24:** Sketch of a typical Cherenkov detector. M = mirror, P = charged particle, PM = photomultiplier.

### *4.6.1 Calibration and sensitivity*

The photon yield dN/dx is defined for the Cherenkov effect:

$$\text{photon yield}: \frac{dN}{dx} = 2\cdot\pi\cdot\alpha\cdot\sin^2\Theta\cdot\left(\frac{1}{\lambda_1}-\frac{1}{\lambda_2}\right)$$

with: $\cos\Theta = 1/(\beta\cdot n)$, $\beta > 1/n$, $\alpha = 1/137.036$ and $\lambda_{1,2}$ = wavelength interval.

A typical photocathode is sensitive between $\lambda_1 = 350 \le \lambda \le \lambda_2 = 500$ nm. Therefore:

$$\frac{dN}{dx} = 390\cdot\sin^2\Theta\,\frac{photons}{cm}$$

As pointed out in the beginning, let us assume that the radiation is created by MIPs which have $\beta = 1$. Fused silica has a refraction index of $n = 1.55$, therefore $\Theta = 49.80$ and

dN/dx = 227 photons/(MIP cm)

A 1 ltr. Cherenkov detector ($10\cdot10\cdot10$ cm$^3$) produces [(Eq. (2))]:

$$\frac{3.1\cdot10^7\,MIP}{cm^2\cdot rad}\cdot\frac{227\,photons}{MIP\cdot cm}\cdot1000\,cm^3 = 7\cdot10^{12}\,\frac{photons}{rad}$$

The collection efficiency of the Cherenkov light is in the order of $\varepsilon_{coll} = 80$ %, which is valid only for directed light created by particles crossing the detector with a common and know angle. This might not be true for beam loss monitoring (=> need Monte Carlo simulation). The collected light is then converted into photoelectrons at the photocathode of the PMT with an efficiency of about $\varepsilon_{cath} = 30$ %. The sensitivity $S_{Che}$ for a 1 ltr. Cherenkov detector connected to a PMT can be calculated by:

$$S_{Che} = 7\cdot10^{12}\,\frac{photons}{rad}\cdot\varepsilon_{coll}\cdot\varepsilon_{cath}\cdot\frac{1.6\cdot10^{-19}\,C}{e^-}\cdot PMT_{gain} = 270\,\frac{nC}{rad}\cdot PMT_{gain}$$

This is only a factor 2–3 smaller than a 1 ltr. ionization chamber but the response time of a Cherenkov detector is much faster. It is like a scintillator counter defined by the response time of the PMT of < 10 ns. But note again that the calculated sensitivity $S_{Che}$ is valid only for directed light owing to the 80% collection efficiency.

Since a very efficient Cherenkov detector needs good mirrors, it might become expensive and its maintenance might become costly. A very cheap Cherenkov BLM was developed at CEBAF using only a simple and cheap PMT [84]. The Cherenkov light is created in the glass tube of the PMT which is then directly detected. The housing of the PMT was made of simple water pipes (see Fig. 25). It is a quite radiation tolerant system; however, the darkening of the PMT glass has to be compensated by increasing the PMT gain. Such a system is not sensitive enough to measure 'small normal' losses but it was used to control and limit strong and dangerous losses.

The stabilization of the PMT gain, as discussed in Section 4.5.1, is also valid in the case of Cherenkov detectors.

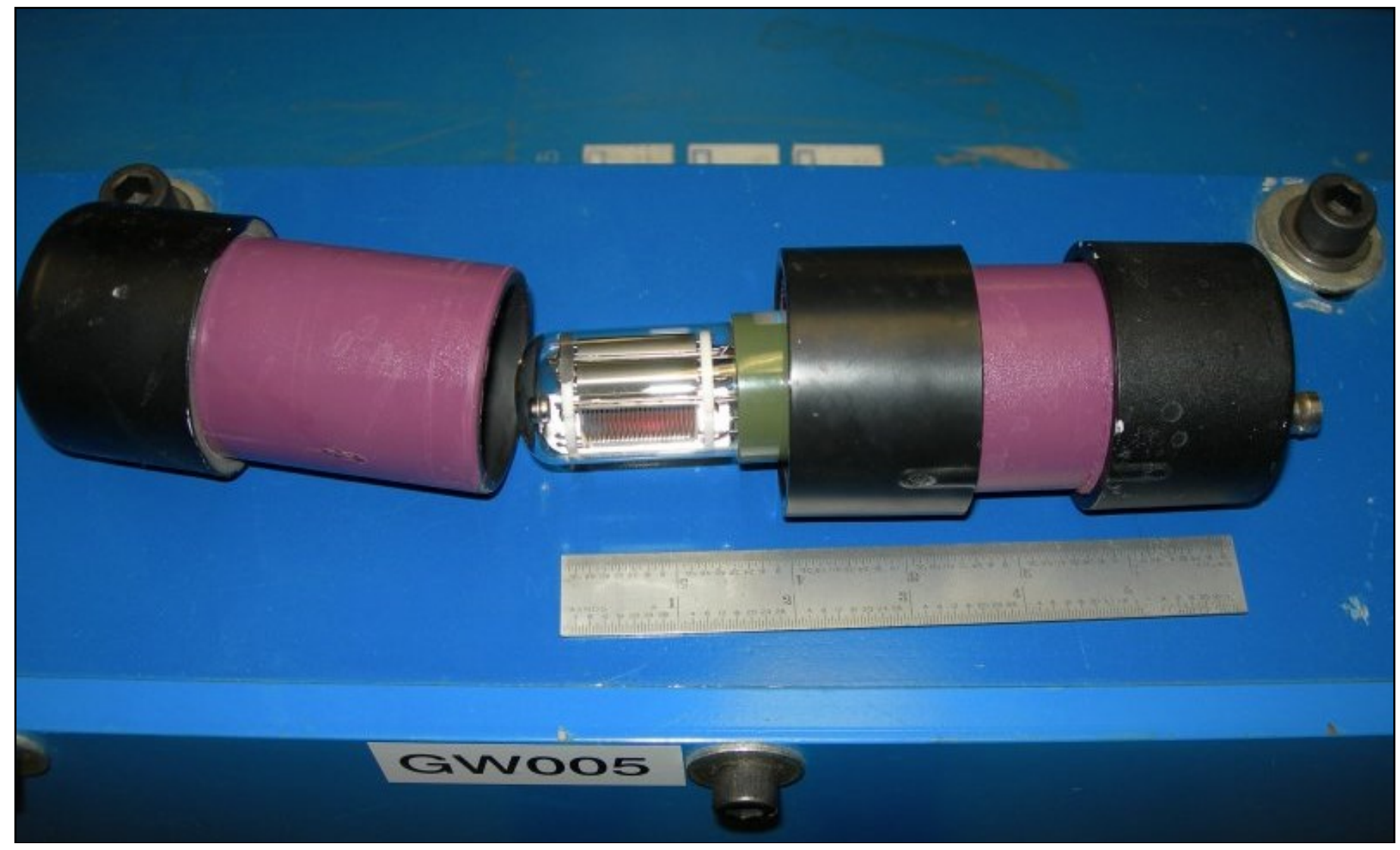

**Fig. 25:** The simple CEBAF Cherenkov BLMs. Courtesy K. Jordan, JLab.

### *4.6.2 Optical fibres*

Another idea for reducing the costs of a BLM system is to reduce the quantity of BLMs but with sufficient covering of the accelerator. One can use Cherenkov light created in long optical fibres to determine the position of beam losses. It allows real-time monitoring of loss location and loss intensity like in PLICs (Section 4.2). The fast response of the Cherenkov signal is detected with photomultipliers at the ends of the irradiated fibres. A time measurement provides the position measurement along the fibre while the integrated light gives the amount of losses (see Fig. 26). A longitudinal position resolution of 20 cm (= 1 ns at $v = 0.66c$) is possible. The dispersion in the fiber of 0.1 - 0.2 ns/m will reduce the position resolution in long fibers. A readout along the beam direction will smear out the resolution [85].

The use of high purity quartz fibres (Suprasil) has some advantages:

- only Cherenkov emission, no scintillation;
- $n$=1.457, high refraction index;
- withstand $30 \cdot 10^9$ rad, radiation hard.

Scintillating fibres give about 1000 times more light output but are much more sensitive to radiation damage (reduction of 1/e at $\approx 10^8$ rad) [86].

There are two major issues to address when considering the sensitivity of BLMs using the Cherenkov effect in single quartz optical fibres.

- The light yield caused by the passage of a single charged particle in a fibre.
- The probability of survival of the emitted photons.

The Cherenkov cone for $\beta = 1$ particles (MIPs) is $\Theta \approx 47^0$.

The condition for light capture and transport down the fibre core is given by:

$$\xi \geq \arcsin\left(\frac{n_{clad}}{n_{core}}\right),$$

where ξ depends on (shower) particle trajectory. $n_{clad}$ is the (lower) refraction index of the cladding which causes the light to be confined to the core of the fibre by total internal reflection at the boundary between cladding and core. The numerical aperture NA is defined as:

$$\sqrt{n_{core}^2 - n_{clad}^2} = NA \approx 0.3 .$$

The distribution of photons trapped inside a fibre is a function of the impacting particle's angle *a* and impact parameter *b*. For more details see Figs. 27, 28 and Ref. [87]. Since the efficiency of Cherenkov fibres depends so strongly on the impact distance and angle, a general formula for the sensitivity of Cherenkov fibre BLMs cannot be given here. Monte Carlo simulations are necessary to calculate the response at certain conditions. However, a very rough estimate can be done by assuming a $L$=100 cm long fibre with a diameter of $R$=0.01 cm which gives a volume of $V=R^2 \cdot \pi \cdot L = 0.03$ cm$^3$. By comparing it with the 1 ltr. Cherenkov detector Signal $S_{che}$ of chap. 4.6.1 and including the numerical aperture $NA$= 0.3 gives a Signal $S_{fibre}$ of:

$$S_{fibre} = S_{Che} \cdot 3 \cdot 10^{-5} \cdot 0.3 \approx 3 \cdot 10^{-3} nC/rad$$

Note that for longer fibres the signal undergoes attenuation during the propagation of the light, about 15% for a 10 m long fibre and more than 60% for a 100 m long fibre, depending on the detection wavelengths [88].

Examples for Cherenkov fibre based BLM systems can be found in Ref. [89]. Also, dose measurements can be done by optical fibres, but they are not treated here because of their long-time response of hours or days [90].

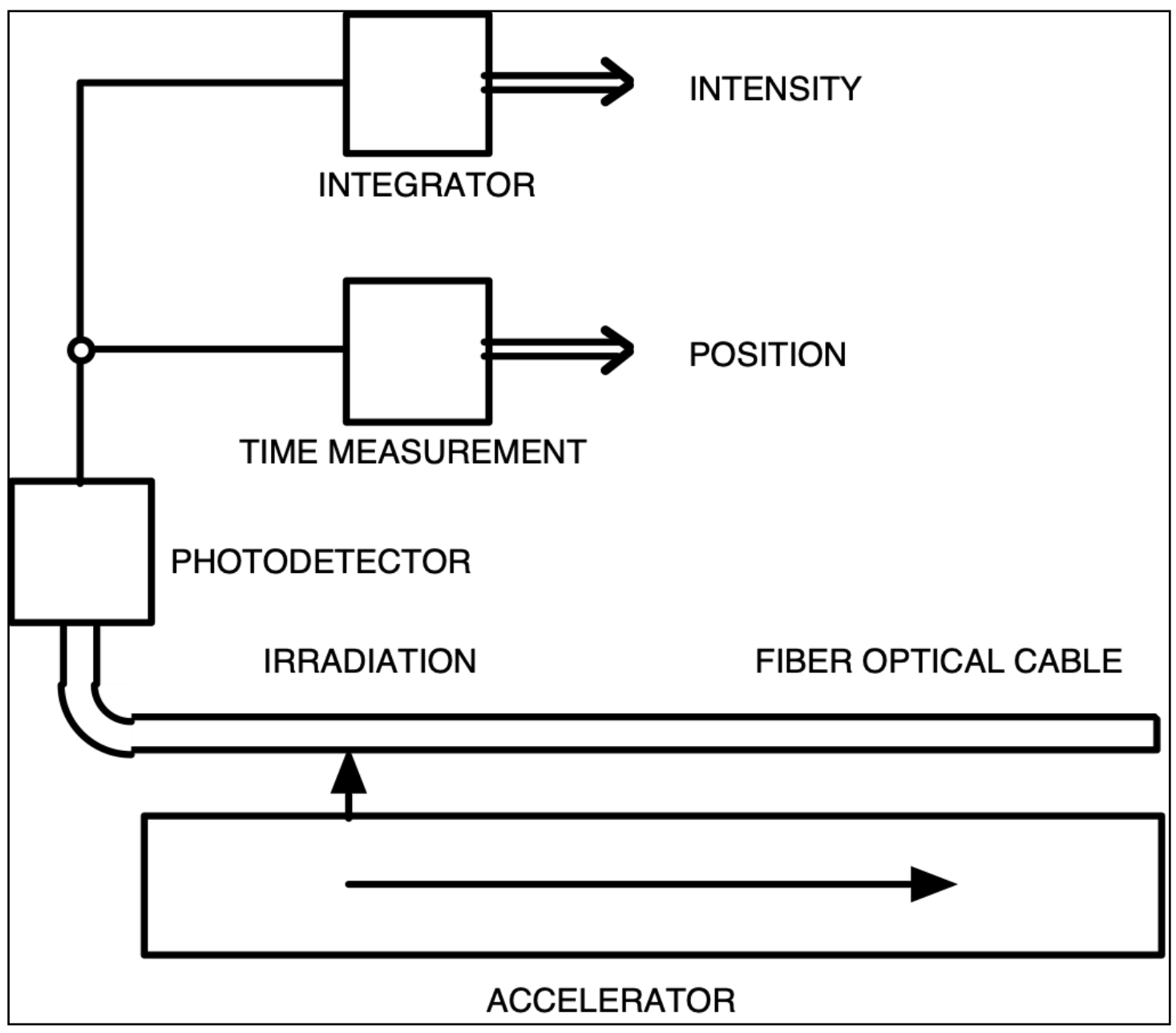

**Fig. 26:** Schematic of the detection of irradiation by means of an optical fibre; from Ref. [91]

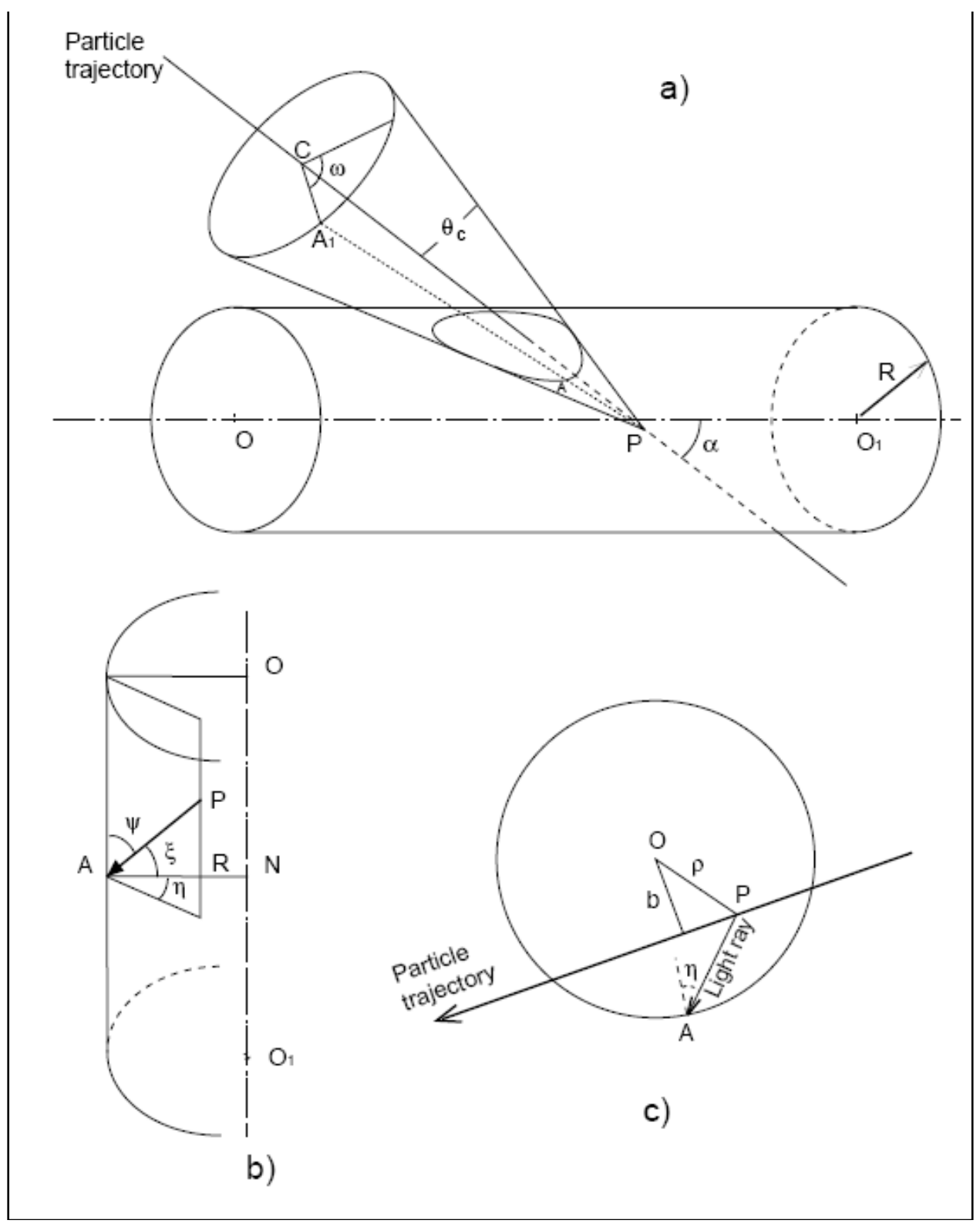

**Fig. 27:** A schematic view of the Cherenkov Effect in a step index optical fibre. A photon emitted at the point P travels along the Cherenkov cone and hits the interface core-cladding at point A. (a) Three-dimensional view; (b) Blow-up view of the point of impact of a light ray on the interface core-cladding; (c) Projection into a plane perpendicular to the fibre axis. From Ref. [87].

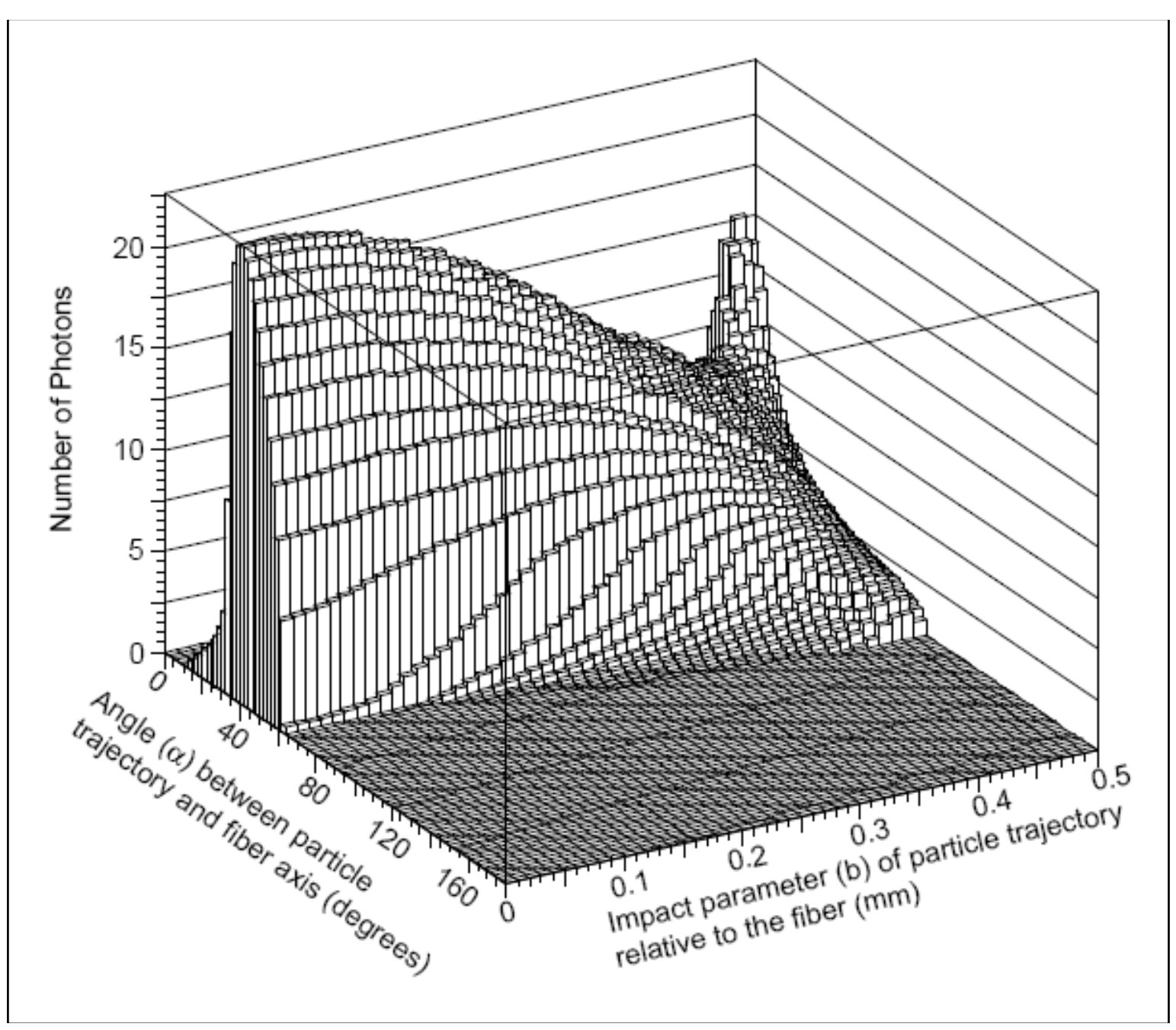

**Fig. 28:** The distribution of photons trapped inside a fibre as a function of the impacting particle's angle and impact parameter b. *NA* = 0.37. From Ref. [87].

### 4.7 Neutron Detectors

The neutron-sensitive detectors (ND) are useful in low energy part of hadron accelerators since the beam loss signal comes only from neutral particles (neutrons and photons) which are almost the only ones able to escape from the beam pipes (no MIPs) [32]. A typical compact neutron detector consists of a neutron converter ($^{6}$Li(n,$\alpha$), $^{10}$B(n,$\alpha$)) together with or in a scintillator (to detect the alphas). The scintillator is connected to a photomultiplier tube surrounded by polyethylene neutron moderator[4] which increase the cross section for capturing thermal neutrons. The converter, scintillator and the PMT are often shielded by some mm of lead to remove x-ray background. Neutron detectors are good at detecting losses occurred meters away from the detector itself since especially low energy neutrons are bounce on the material of the tunnel etc. over many meters. A conversion of such a lost signal back into number of lost particles is only possible by detailed Monte Carlo studies. There is no simple "conversion factor", since the same loss reading could correspond to many different loss scenarios [32]. Slow neutron detectors respond to any loss and are effectively covering a wider region but are incapable of spatial resolution. At SNS a combination of slow neutron detectors and ionization chambers are used for sensitive beam loss detection [92], [93].

An additional issue was introduced at SNS and at ESS by high x-ray noise originating from the superconducting linac (SCL) cavity dark current. This radiation might give a strong signal in the neutron detectors which might become larger than the signal form beam losses. There are some commercial detector systems available with some rejection of $\gamma$-background based on diamonds [94], [95], scintillators [96] and proportional counter tubes [97], just as a few examples. A new BLM system based on fast neutron detection was recently developed for ESS [98]. It consist of micromegas detectors [99], [100] sensitive to fast neutrons and "blind" to thermal neutrons by shielding and "blind" to X- and $\gamma$-rays based on signal discrimination. For a fast version of the detector, with a response time of ns, a detection efficiency of $10^{-5}$ to $10^{-3}$ for neutron energies 0.5-10 MeV was reported. A 25 x 25 cm$^2$ detector is expected to give a count rate of few counts/s for a neutron fluence rate of 1 neutron/(s cm$^2$). A 1 rad dose on neutrons (1 MeV) in soft tissue is produce by $4 \cdot 10^{8}$ n/cm$^2$ [101]. That gives a sensitivity of roughly some nrad/count. A slow version of the detector with a polyethylene moderator reached a few % efficiency but with time delays up to 200 μs. Many more details on this BLM System can be found in [102] and [103].

## 5 Summary

Different types of beam losses together with some examples were shown. Beam loss monitoring techniques for measuring losses along an entire accelerator have been discussed with the focus on the sensitivity of the various types.

The most common BLM is a short ionization chamber. Whether a simple air-filled chamber is adequate, or an argon or helium filled chamber with superior higher dynamic range must be used, depends on the conditions of the particular accelerator. Ionization chambers are radiation resistant but respond to synchrotron radiation.

Long ionization chambers using a single coaxial cable work well for one-shot accelerators or transport lines. To achieve spatial resolution of losses along an entire accelerator two conditions must be fulfilled: 1) The machine must be much longer than the bunch train and 2) the particles must be relativistic, or 3) the long chamber has to be split into short parts which are read out individually.

PIN diodes and CVD diamonds with thick depletion layers can be used as 'solid state' ionization chambers. They have a high sensitivity, but they exist only in small sizes. The combination of two PIN photodiodes in a coincidence counting mode results in a detector with very large dynamic range and extremely effective rejection of synchrotron radiation. The measured radiation resistance permits long term use also in high energy electron machines with a high radiation background. A limitation is the inability to distinguish overlapping counts, so that the response is linear only for losses which are less than one count per coincidence interval.

[4] Gas proportional counters neutron detectors use Helium (3He(n,p)) or BF3 (10B(n, a) interactions.

A very sensitive system for measuring beam losses is a PMT in combination with a scintillator. Because of the adjustable gain the dynamic range can be large, but the calibration of each device must be adjusted and monitored over time. These systems are also sensitive to synchrotron radiation and are relatively expensive.

Cherenkov counters are insensitive to synchrotron radiation background, but they are less sensitive to beam losses. However, with the additional gain of a PMT their sensitivity exceeds the ionization chamber. Long optical fibres can be used like long ionization chambers with the same limitations in the bunch repetition rate. Cherenkov based fibres are much more radiation hard but much less sensitive to losses than scintillating fibres.

The following table summarizes the estimated sensitivity of the BLM types discussed together with some interesting numbers. Note that the estimated sensitivity is valid for doses created by MIPs only, therefore no sensitivities of neutron detectors are given. Careful Monte Carlo studies are necessary to get the real response of a BLM at its location. A detailed understanding of the reasons for beam losses is most helpful to get the right starting point for the simulation, e.g., where a beam loss might occur.

| **Detector Material** | **Energy to create one electron [eV/e]** | **Number of [e / (cm MIP)] (depends on dE/dx, resp. density)** | **Sensitivity S (for MIPs) [nC/rad]** |
|---|---|---|---|
| Plastic Scintillator | 250-2500 | $10^3 - 10^4$ | ≈18·103 (· PMTgain)<br>(1 ltr.) |
| Inorganic Scint. | 50-250 | $10^4 - 10^5$ | ≈ 200·103 (· PMTgain)<br>(1 ltr.) |
| Gas Ionization | 22-95 | ≈100<br>(Ar,1 atm., 200C) | ≈ 600 (· Elecgain)<br>(1ltr) |
| Semiconductor (Si) | 3.6 | $10^6$ | ≈ 100 (· Elecgain)<br>(1 cm2 PIN-Diode) |
| CVD Diamond | 13 | $4 \cdot 10^5$ | ≈ 44 (· Elecgain)<br>(1 cm2 Diamond) |
| Secondary emission | 2-5%/MIP<br>(surface only) | 0.02-0.05 e/MIP | ≈ 2·10-3 (· PMTgain)<br>(8cm2) |
| Cherenkov light | $10^5 - 10^6$ | ≈10 ($H_2O$) | ≈ 270 (· PMTgain)<br>(1 ltr.) |
| Quartz fibre | | | ≈ 3·10-3 (· PMTgain)<br>(1 m) |

**Table 1:** The estimated sensitivity of the BLM types

The different detectors span a huge area of sensitivities, $S > 10^8$. This offers an extended dynamic range of measuring beam losses just by using different types of detectors. Ref. [104] uses this technique to extend the useful dynamic range and the speed of loss measurements by using air filled ionization chambers + scintillators + proportional chambers archiving a dynamic range of 160 dB.

## Acknowledgements

Many thanks to my colleagues Mark Lomperski, Lars Fröhlich, and Gero Kube from DESY for adding a lot of useful hints and comments to this manuscript. Many authors listed in the references provided me with their talks, pictures or data, which is gratefully acknowledged.